\documentclass[aps,prd,twocolumn,showkeys,superscriptaddress,nofootinbib,floatfix]{revtex4-2}
\usepackage[utf8]{inputenc}
\usepackage[T1]{fontenc}
\usepackage{amsmath}
\usepackage{amssymb}
\usepackage{bm}
\usepackage{graphicx}       
\usepackage{subcaption}     
\usepackage{float}          
\usepackage{xcolor}       
\usepackage{eurosym}      
\usepackage{newunicodechar}
\begin{document}

\title{Breakdown of Smooth Shock Solutions in Transient Relativistic
Hydrodynamics}
\author{Davi D. Oliveira}
\email{davioliveira@id.uff.br}
\affiliation{Instituto de Física, Universidade Federal Fluminense, Niterói, Rio de Janeiro, 24210-346, Brazil}
\author{Gabriel S. Denicol}
\email{gsdenicol@id.uff.br}
\affiliation{Instituto de Física, Universidade Federal Fluminense, Niterói, Rio de Janeiro, 24210-346, Brazil}

\begin{abstract}
In this work, we demonstrate that shock solutions in the Israel-Stewart framework lose regularity once the shock velocity reaches a critical value, and a discontinuity emerges in the solution, which can be interpreted as a second shock wave. This subshock arises as a consequence of the finite speed of information propagation inherent to the Israel-Stewart theory. We then perform numerical simulations to confirm the breakdown of solution continuity. Subsequently, we propose two regularization procedures to extend the domain of continuous shock solutions. The first employs a third-order extension, which introduces new kinetic fields, while the second incorporates a small numerical bulk viscosity. Both methods effectively increase the maximum propagation speed of the Israel-Stewart theory and extend the range over which regular shock solutions exist. Thus, we confirm that this loss of regularity is a direct consequence of the Israel-Stewart framework. These results suggest that Israel-Stewart theory may not provide an adequate description of ultra-relativistic shock waves.
\end{abstract}

\maketitle

\section{Introduction}

Shock waves are among the most fundamental nonlinear structures in fluid dynamics, providing a paradigmatic example of how hydrodynamic theories behave in the presence of large gradients and strong departures from equilibrium \cite{Landau}.
They appear in a wide range of physical systems, from
astrophysical phenomena such as supernova explosions \cite{Gulliford1974}
and neutron star mergers \cite{verma2022importance} to the strongly
interacting matter produced in relativistic heavy-ion-collisions \cite
{Scheid_PhysRevLett.32.741,RISCHKE1996479,Bouras_2009_0902.1927,H.H_Gutbrod_PhysRevC.42.640}. 

In ideal relativistic hydrodynamics, shock waves are described as
discontinuities satisfying the Rankine-Hugoniot conditions \cite{Landau}, first derived in
the relativistic context by Taub \cite{Taub1973-cf}. The inclusion of
dissipation is generally expected to regularize these discontinuities into
smooth profiles with a finite thickness \cite{Landau}. In the relativistic regime, including dissipation can be nontrivial since the traditional Navier-Stokes theory is known
to be acausal and unstable \cite{PhysRevD.31.725}. This led to the
development of causal theories of dissipative hydrodynamics, most notably
the Israel-Stewart (IS) framework \cite{ISRAEL1976310,ISRAEL1979}, in which dissipative
currents relax on finite timescales. While IS theory restores causality
under appropriate conditions, it also introduces a finite characteristic
propagation speed, which can play a central role in
the nonlinear dynamics of the system.

Relativistic shock waves in dissipative hydrodynamics have been extensively
investigated in the context of heavy-ion collisions, particularly through
studies of the relativistic Riemann problem \cite%
{Riemann_loraclavijo2013exact}. A series of works by the Frankfurt group \cite{Rischke_1995}, including detailed comparisons with
kinetic theory based on the Boltzmann equation \cite%
{Bouras_2009_0902.1927,Moln_r_2009,Denicol_PhysRevD.85.114047}, provided important insights into the structure of relativistic viscous shock waves and their description within Israel-Stewart theory.  In practice, however, most analyses have focused on weak or moderately strong shocks, and the structure of ultra-relativistic shocks in causal dissipative theories remains comparatively unexplored. In this regime, a qualitatively new situation may arise, since the shock velocity can become comparable to the characteristic propagation speeds supported by the hydrodynamic theory.

The implications of this regime for the stability of stationary shock solutions has been initially explored by Calzetta \cite{Calzetta_2022}, who showed, through a linearized analysis, that the decay rate of stationary shock profiles diverges as the flow velocity approaches the maximum propagation speed supported by the theory. More recently, Bemfica \cite{Bemfica_2025} demonstrated that smooth initial data in Israel-Stewart theory can develop finite-time gradient singularities, providing a dynamical mechanism for shock formation in the nonlinear regime. These developments build upon a broader understanding of causality and characteristic propagation in relativistic hydrodynamics developed by Bemfica, Disconzi, Noronha and collaborators, who derived conditions for nonlinear causality and local well-posedness while clarifying the role of the characteristic structure of the equations \cite{Bemfica_2018,Bemfica:2020xym}. Taken together, these developments suggest that characteristic propagation speeds play a fundamental role in the dynamics of relativistic shock waves.

The possibility that characteristic propagation speeds may constrain the structure of shock waves has a well-known analogue in extended thermodynamics \cite{Micenmacher1989-fx,Jou1991-od}. There, smooth shock profiles cease to exist above a critical Mach number and are replaced by composite structures containing internal discontinuities, commonly referred to as subshocks \cite{MTorri_shocktube,MTorri_13momento}. These structures arise when the characteristic propagation speeds supported by a given moment closure become insufficient to continuously resolve the shock profile. As we shall show, an analogous mechanism operates in Israel-Stewart theory.

These developments naturally raise the question of whether the characteristic propagation speeds of causal relativistic hydrodynamics also constrain the existence of smooth stationary shock solutions. In this work we address this question by investigating stationary shock waves in conformal Israel-Stewart theory in the ultra-relativistic regime. We derive the conditions under which smooth stationary shock profiles exist and show that they break down once the shock velocity reaches the characteristic propagation speed of the nonlinear hydrodynamic equations. Beyond this point, the stationary equations lose regularity and the shock must be completed by a discontinuity satisfying the Rankine-Hugoniot conditions, yielding a relativistic analogue of the subshocks encountered in extended thermodynamics. We confirm this picture through numerical solutions of the relativistic Riemann problem, where the embedded discontinuity emerges dynamically from the full time-dependent evolution. Finally, we investigate two regularization strategies—a third-order extension derived from kinetic theory and a numerical bulk-viscosity regulator—and show that both restore smooth solutions by increasing the characteristic propagation speed of the theory.

The paper is organized as follows. Section \ref{SecII} reviews the relevant aspects of relativistic dissipative hydrodynamics. In Sec.~\ref{Sec:ShockWave} we derive the stationary shock equations and establish the criterion for the breakdown of smooth solutions. Section \ref{SecIV} compares these analytical predictions with numerical solutions of the relativistic Riemann problem. Sections \ref{SecV} and \ref{SecVI} present two regularization strategies based on a third-order extension of the theory and on an effective numerical bulk viscosity, respectively. We then end with our conclusions.

\textbf{Notation:} We use natural units $\hbar = c = k_{B} = 1$ and
the mostly minus $(+,-,-,-)$ metric signature.

\section{Relativistic hydrodynamics}
\label{SecII}
In this section, we introduce the relativistic hydrodynamic framework used throughout this work, emphasizing the aspects relevant to the analysis of stationary shock waves.

\subsection{Basic equations}

We consider a conformal relativistic fluid without conserved charges, whose dynamics is governed by the local conservation of energy and momentum \cite{Landau}, 
\begin{equation}
\partial _{\mu }T^{\mu \nu }=0.  \label{conserv}
\end{equation}
The energy-momentum tensor is decomposed in terms of a normalized 4-velocity, $u^{\mu }$, as
\begin{equation}
T^{\mu \nu }=\varepsilon u^{\mu }u^{\nu }-p\Delta^{\mu\nu
}+\pi ^{\mu \nu },  \label{TensorEnergiaMomento}
\end{equation}
where $\varepsilon $ is the energy density, $p$ the thermodynamic pressure, $\Delta^{\mu\nu}=g^{\mu\nu}-u^\mu u^\nu$ is the projection operator orthogonal to $u^\mu$, and $\pi ^{\mu \nu }$ the shear-stress tensor. The pressure is given by the
equation of state of a conformal fluid \cite{Baier_2008},  
\begin{equation}
p=\frac{1}{3}\varepsilon .  \label{TensorEnergiaMomento}
\end{equation}%

Dissipative effects are described within Israel-Stewart theory \cite{ISRAEL1976310,ISRAEL1979}, in which the shear-stress tensor obeys a relaxation-type equation, 
\begin{equation}
\tau _{\pi }\Delta _{\alpha }^{\mu} \Delta _{\beta }^{ \nu }u^{\lambda }\partial _{\lambda
}\pi ^{\alpha \beta }+\pi ^{\mu \nu }=2\eta \sigma ^{\mu \nu }-\frac{4}{3}%
\tau _{\pi }\pi ^{\mu \nu }\partial _{\lambda }u^{\lambda } + \cdots,
\label{IsraelStewartMusic}
\end{equation}%
where $\eta$ is the shear viscosity and $\tau _{\pi }$ is the shear relaxation time. The ellipsis denote possible higher-order terms that will not be considered in this work. We also introduced the shear tensor, 
\begin{equation}
\sigma^{\mu \nu } \equiv \frac{1}{2}\left(\nabla^\mu u^\nu + \nabla^\nu u^\mu\right)-\frac{1}{3}\Delta^{\mu\nu}\partial_\lambda u^\lambda,
\end{equation}
with $\nabla^\mu \equiv \Delta^{\mu\nu} \partial_\nu$ being the space-like gradient. We note that the Navier-Stokes limit is recovered for $\tau_\pi \rightarrow 0$, in which case $\pi^{\mu \nu }=2\eta\sigma^{\mu \nu }$. For simplicity, we restrict our analysis to this minimal version of Israel-Stewart theory and neglect the remaining second-order terms and transport coefficients. The omitted terms do not modify the qualitative features of the stationary shock solutions discussed below.

\subsection{Causality and maximum propagation speed}

Relativistic Navier-Stokes theory is known to be acausal, as dissipative currents respond instantaneously to gradients, leading to arbitrarily fast signal propagation. Israel-Stewart theory resolves this issue by introducing
finite relaxation times, which render the equations hyperbolic in the linear regime. A direct consequence of this structure is the existence of a finite characteristic propagation speed for disturbances in the fluid. Linearizing the equations around equilibrium and analyzing the large-wavenumber limit of the dispersion relation, one finds the asymptotic characteristic velocity \cite{Brito_2020}, 
\begin{equation}
v_{\mathrm{L}}^{2}=\frac{1}{3}+\frac{4\eta }{3\tau _{\pi }(\varepsilon +p)}.
\end{equation}%
Physically, $v_{\mathrm{L}}$ sets the maximum speed at which information can propagate within the linearized hydrodynamic description. Naturally, causality requires $v_{\mathrm{L}}\leq 1$, which imposes constraints on the
transport coefficients -- in particular, it prohibits the relaxation time to be taken to zero and provides a lower bound for this quantity.

However, shock waves probe the fully nonlinear regime, where additional constraints on causality arise. Nonlinear causality conditions for relativistic dissipative hydrodynamics have been derived by Bemfica, Disconzi and Noronha in Ref.~\cite{Bemfica:2020xym} and recently complemented in Ref.~\cite{cordeiro2026nonlinearcausalitystronghyperbolicity}. In the absence of conserved charges and considering only one-dimensional evolutions, one obtains the following maximum propagation speed, 
\begin{equation}
v_{\mathrm{NL}}^{2}=\frac{1}{3}+\frac{4}{3}\frac{\eta +\tau _{\pi }\pi }{%
\tau _{\pi }(\varepsilon +p+\pi )}\leq 1,  \label{causality Benfica 1}
\end{equation}%
with $\pi ^{xx}=\gamma ^{2}\pi $. In contrast to the linear condition, this bound depends explicitly on the local value of the dissipative stress and therefore becomes nontrivial far from equilibrium. In particular, large values of $\pi $, such as those that may arise near shock fronts, can drive
the system toward the boundary of its causal domain even if the transport coefficients satisfy the linear constraint. This highlights that causality in relativistic dissipative hydrodynamics is not only a property of the transport coefficients, but also of the dynamical state of the system. We note that the nonlinear condition reduces to the linear one in the limit of $\pi \rightarrow 0$, as expected. In the presence of shear viscosity, this
bound becomes more complicated, but can be found in Ref. \cite{Bemfica_2018}.

As we will show in the following section, the characteristic propagation speed provides the central criterion controlling the existence of smooth stationary shock profiles.

\section{Shock Wave}
\label{Sec:ShockWave}

\subsection{Stationary solutions}

Shock waves are nonlinear, supersonic disturbances that generate rapid variations of the hydrodynamic fields. 

In this section, we derive these solutions within the Israel-Stewart framework introduced in Sec.~\ref{SecII}, with the aim of identifying the conditions under which smooth shock profiles can exist. We restrict to one-dimensional configurations and consider traveling wave solutions in which all hydrodynamic fields depend on the variable,
\begin{equation}
\omega=x-v_{\text{shock}}t,
\end{equation} where $v_{\text{shock}}$ is the shock velocity. In the rest frame of the shock ($v_{\text{shock}}=0$), the fields depend only
on a single spatial coordinate $\omega $ and can be described as stationary solutions of the full nonlinear hydrodynamic equations. In this case, the conservation laws \eqref{conserv} reduce to, 
\begin{eqnarray}
\partial _{\omega }[(\varepsilon +p+\pi )\gamma ^{2}v] &=&0,  \label{Bla1} \\
\partial _{\omega }[(\varepsilon +p+\pi )\gamma ^{2}v^{2}+p+\pi ] &=&0,
\label{Bla2}
\end{eqnarray}
where $\gamma =1/\sqrt{1-v^{2}}$ is the Lorentz factor and $v$ is the fluid velocity in the rest frame of the shock. For the sake of simplicity, we further introduced the variable,
\begin{equation*}
\pi \equiv \frac{\pi ^{xx}}{\gamma ^{2}}.
\end{equation*}%
In this form, the shear contribution to the conserved current has the identical structure of bulk viscosity contributions. The equation of motion
for $\pi $ reduces to, 
\begin{equation}
\gamma v\partial _{\omega }\pi +\frac{\pi }{\tau _{\pi}}=-\frac{4}{3}\left( 
\frac{\eta }{\tau _{\pi}}+\pi \right) \gamma ^{3}\partial _{\omega }v.
\label{ShearEquation}
\end{equation}

Equations \eqref{Bla1} and \eqref{Bla2} are solved by, 
\begin{eqnarray}
(\varepsilon +p+\pi )\gamma ^{2}v &=&C_{1},  \label{Constante1} \\
(\varepsilon +p+\pi )\gamma ^{2}v^{2}+p+\pi  &=&C_{2},  \label{Constante2}
\end{eqnarray}%
where $C_{1}$ and $C_{2}$ are constants. These equations can be reduced to the following simple form, 
\begin{equation}
-C_{1}v^{2}+\frac{4}{3}C_{2}v-\frac{C_{1}}{3}=\pi v.
\label{EquacaoChoqueGrande}
\end{equation}
This is one of the central  equations in this context of shock solutions in hydrodynamics. In the ideal fluid limit, $\pi =0$, this equation reduces to an algebraic quadratic equation for the velocity field, 
\begin{equation}
\frac{C_{1}}{C_{2}}(3v^{2}+1)-4v=0,
\end{equation}%
which admits two constant solutions, 
\begin{equation}
v_{i/f}=\frac{2C_{2}}{3C_{1}}\pm \sqrt{\left( \frac{2C_{2}}{3C_{1}}\right)
^{2}-\frac{1}{3}}.  \label{v inicial e v final}
\end{equation}
Since the equation is purely algebraic, no continuous trajectory can connect $v_{i}$ and $v_{f}$. The physically relevant solution therefore consists of two constant states separated by a discontinuity, with $v_{i}$ and $v_{f}$ representing the asymptotic fluid velocities on either side of the shock. This is precisely the standard Rankine-Hugoniot shock solution of ideal hydrodynamics. Here we choose $C_{1}<0,$ such that both asymptotic velocities are negative and satisfy $\left\vert v_{i}\right\vert
<\left\vert v_{f}\right\vert$. Therefore, in this frame the fluid flows from the state with velocity $v_f$ toward the state with velocity $v_i$ and we identify $v_f$ and $v_i$ as upstream and downstream velocities, respectively. Conservation of the energy-momentum flux further implies that the corresponding energy densities of the asymptotic states satisfy, $\varepsilon_{i}>\varepsilon _{f}$. Finally, we note that in the rest frame of the shock, the downstream velocity can be identified with the shock velocity in the laboratory frame, via $v_{f}=-v_{\text{shock}}$.

The inclusion of dissipation is expected to regularize the aforementioned discontinuity, turning the original algebraic equation for the velocity fields into ordinary differential equations. To illustrate this, we first
consider the Navier-Stokes limit, obtained from Eq.~\eqref{ShearEquation} by taking $\tau _{\pi}\rightarrow 0$, which yields the constitutive relation 
\begin{equation}
\pi =-\frac{4}{3}\eta \gamma ^{3}\partial _{\omega }v.  \label{Constitutve}
\end{equation}%
If we substitute it in Eq. \eqref{EquacaoChoqueGrande} and rewrite the left hand side in terms of $v_{i}$ and $v_{f}$, we obtain the following first-order differential equation for the velocity field, 
\begin{equation}
C_{1}\frac{\left( v-v_{i}\right) \left( v-v_{f}\right) }{v}=\frac{4}{3}\eta
\gamma ^{3}\partial _{\omega }v.  \label{Fundamental}
\end{equation}%
In contrast to the ideal case, this equation admits smooth solutions in which
the velocity interpolates continuously between $v_{i}$ and $v_{f}$ \cite{Landau}, with the shear viscosity controlling the thickness of the resulting shock profile. This equation shows that if the velocity profile starts within the range $\left\vert v_{i}\right\vert <\left\vert v\left( \omega
\right) \right\vert <\left\vert v_{f}\right\vert $, the velocity gradient will always be negative, $\partial_{\omega }v<0$, leading to a monotonic velocity field.

We now turn to Israel-Stewart theory, in which the dissipative current satisfies an independent relaxation equation. Combining Eq.~\eqref{EquacaoChoqueGrande} with Eq.~\eqref{ShearEquation}, we rewrite the conservation laws in a form that isolates $\partial_{\omega }v$, leading to
a first-order differential equation for the velocity that remains coupled to the dissipative current,
\begin{equation}
\left( \varepsilon +p+\pi \right) \left( v^{2}-v_{\mathrm{NL}}^{2}\right)
\gamma ^{3}\partial _{x}v=\frac{\pi }{\tau _{\pi }},  \label{MasterEquation}
\end{equation}
where we identify the maximum propagation speed%
\begin{equation}
v_{\mathrm{NL}}^{2}=\frac{1}{3}+\frac{1}{\varepsilon +p+\pi }\frac{4}{3}\left( \frac{\eta }{\tau _{\pi }}+\pi \right) .
\label{eq:V_NL}
\end{equation}
Here we have used the relations between the integration constants and that the asymptotic velocities satisfy, $v_{i}v_{f}=1/3$. Physically, the shock is still expected to have a finite width, with the system approaching
equilibrium far from the shock front. Accordingly, the velocity profile should interpolate smoothly between $v_{i}$ and $v_{f}$ as $\omega $ varies from $\infty $ to $-\infty $. We thus search for solutions that satisfy the boundary conditions,
\begin{equation*}
\lim_{\omega \rightarrow \infty }v\left( \omega \right) =v_{i}\text{ and } \lim_{\omega \rightarrow -\infty }v\left( \omega \right) =v_{f}.
\end{equation*}

A necessary condition for the existence of a smooth solution of Eq.~\eqref{MasterEquation} is that the coefficient multiplying the velocity gradient remains nonzero throughout the profile, i.e., 
\begin{equation}
\left( \varepsilon +p+\pi \right) \left( v^{2}-v_{\mathrm{NL}}^{2}\right)
\neq 0\quad \text{for all }\omega .
\end{equation}%
This condition ensures that the differential equation is well-defined and can be solved for $\partial_{\omega }v$ at every point along the profile. When this condition is violated, the coefficient of the velocity gradient vanishes and the differential equation loses regularity. 

\subsection{Characteristic velocity}
The appearance of the critical velocity $v_{\mathrm{NL}}$ suggests that the breakdown of smooth stationary solutions is closely related to the causal structure of the underlying hydrodynamic theory. To make this connection
explicit, we now briefly discuss the characteristic structure of the dynamical equations and its relation to the traveling-wave reduction employed here.

The characteristic propagation velocities of relativistic hydrodynamics are determined from the principal part of the full nonlinear equations of motion. Schematically, the hydrodynamic equations considered in the previous sections can be written in the form \cite{Bemfica_2018},
\begin{equation}
A^{\mu }\left( U\right) \partial _{\mu }U=S\left( U\right) ,
\end{equation}%
where $U$ denotes the set of dynamical fields, $\left\{ \varepsilon ,v,\pi\right\} $, and $A^{\mu }$ denotes matrices that depend nonlinearly on the local hydrodynamic state. The characteristic hypersurfaces are determined by
the condition \cite{courant_hilbert_1989},
\begin{equation}
\det \left( A^{\mu }\xi _{\mu }\right) =0,
\end{equation}%
where $\xi _{\mu }$ is the normal covector to the characteristic surface. In one spatial dimension, one may write,
\begin{equation}
\xi _{\mu }=\left( -\lambda ,1\right),
\end{equation}%
where $\lambda $ represents the propagation velocity of the disturbance. The characteristic velocities are therefore obtained from \cite{courant_hilbert_1989},
\begin{equation}
\det \left( A^{x}-\lambda A^{t}\right) =0.
\end{equation}%
The roots $\lambda $ determine the local characteristic propagation velocities supported by the nonlinear hydrodynamic theory. In causal relativistic hydrodynamics, all characteristic velocities must remain subluminal  in the local rest frame of the fluid.

The connection with the stationary shock analysis is immediate. For traveling-wave solutions in one dimension, the dynamical fields satisfy,
\begin{equation}
U\left( t,x\right) =U\left( \omega \right) ,\left. {}\right. \omega =x-v_{%
\mathrm{shock}}t,
\end{equation}
and the equations of motion reduce to,%
\begin{equation}
\left( A^{x}-v_{\mathrm{shock}}A^{t}\right) \partial _{\omega }U=S\left(
U\right) .
\end{equation}%
The stationary system therefore loses regularity whenever%
\begin{equation}
~\det \left( A^{x}-v_{\mathrm{shock}}A^{t}\right) =0,
\end{equation}
which coincides with the characteristic condition above. In one spatial dimension, smooth stationary shock solutions therefore cease to exist when the shock velocity reaches one of the characteristic propagation velocities of the nonlinear theory.

\subsection{Conditions for breakdown of smooth solutions}
\label{Sec:breakdown}

We now discuss the explicit conditions for traveling waves in Israel-Stewart theory to break down. We begin this section by demonstrating that Israel-Stewart theory preserves the same sign structure for $\pi $ and $
\partial _{\omega }v$ observed in Navier-Stokes theory. As already mentioned, one expects compressive shocks to satisfy $\partial_\mu u^\mu <0$, which in the Navier-Stokes limit leads to a positive value of $\pi $. In Israel-Stewart theory, the dissipative currents relax toward their corresponding Navier-Stokes values, so one likewise expects $\pi >0$ across the shock profile. We now show that this is indeed the case for the class of stationary solutions considered here.

Consider solutions that satisfy the boundary conditions $\pi(\omega\rightarrow\pm\infty)=0^{+}$ together with $\left\vert
v_{i}\right\vert \leq \left\vert v(\omega)\right\vert \leq \left\vert v_{f}\right\vert$. As long as the stationary equations remain regular, Eq.~\eqref{MasterEquation} determines the sign of the velocity gradient. Since the right-hand side is proportional to $\pi$, positivity of the dissipative current immediately implies,
\begin{equation}
\pi>0
\quad\Longrightarrow\quad
\partial_\omega v<0,
\end{equation}
provided $|v|<|v_{\mathrm{NL}}|$. That is, the positivity of the dissipative component implies that the velocity profile is monotonic across the shock layer.


We now show that the converse statement also holds. Consider a solution satisfying the asymptotic condition $\pi(\omega\to\pm\infty)=0^+$ and assume that the velocity profile remains monotonic, i.e., $\partial_\omega v<0$. If the dissipative current were to become negative at some finite value of $\omega$, the solution would necessarily have to cross $\pi=0$ twice: once when entering the region $\pi<0$ and once when leaving it in order to satisfy the asymptotic boundary conditions. However, evaluating the evolution equation at $\pi=0$ yields $\partial_\omega\pi<0$, as long as the regularity condition remains satisfied. This fixes the local behavior of the solution at every point where $\pi=0$ and is incompatible with the pair of crossings required for the solution to leave and subsequently re-enter the region $\pi>0$. Thus there is no regular solution for $\pi$ satisfying the prescribed boundary conditions that can become negative as long as $\partial_\omega v<0$.

This proves that,
\begin{equation}
\pi >0\Longleftrightarrow \partial _{\omega }v<0,
\end{equation}
as long as the regularity and the asymptotic boundary conditions remain satisfied. An immediate consequence of the monotonicity of the velocity profile is that the maximum value of $v^{2}$ is reached asymptotically upstream, where $\left\vert v\right\vert \rightarrow \left\vert v_{\mathrm{%
shock}}\right\vert $. The shock velocity is thus the largest velocity magnitude attained within the stationary profile.

Since the right-hand side of Eq.~\eqref{MasterEquation} remains positive throughout the regular profile and $\partial_\omega v<0$, the coefficient multiplying the velocity gradient must remain negative,
\begin{equation}
\left( \varepsilon +p+\pi \right) \left( v^{2}-v_{\mathrm{NL}}^{2}\right) <0.
\end{equation}
The monotonicity result derived above now allows this condition to be stated entirely in terms of the asymptotic shock velocity, 
\begin{equation}
v_{\text{shock}}^{2}<v_{\mathrm{NL}}^{2}.
\end{equation}
This constitutes the central condition for the existence of smooth stationary shock profiles in Israel-Stewart theory.
This is not the first indication that the propagation speed imposes a restriction on shock solutions. Calzetta \cite{Calzetta_2022} showed, through a linearized analysis of the asymptotic equilibrium state, that the decay rate of stationary shock profiles diverges as the flow velocity approaches the maximum propagation speed of the linearized theory.

When the asymptotic shock velocity becomes larger than the characteristic propagation speed, the monotonicity of the velocity field guarantees that it must intersect $v_{\mathrm{NL}}$ at some finite value of $\omega$. At this point the coefficient multiplying the velocity gradient vanishes while the numerator remains finite due to the positivity of the dissipative component. The stationary equations therefore become singular and the velocity gradient necessarily diverges at that  point. Smooth continuation of the stationary profile becomes impossible, forcing the solution to develop a discontinuity. The resulting configuration corresponds to a composite shock structure consisting of a smooth viscous precursor followed by an embedded discontinuity. This mechanism is directly analogous to subshock formation in extended thermodynamics, where the loss of regularity of the governing equations necessitates the appearance of an internal discontinuity \cite{MTorri_shocktube}. Beyond the critical velocity, the smooth branch terminates at a characteristic singularity and the shock must be completed by a discontinuity satisfying the Rankine-Hugoniot conditions.

We conclude that smooth viscous shock solutions in Israel-Stewart theory exist only when the shock velocity does not exceed the maximum propagation speed of the theory. Beyond this threshold, the stationary solution necessarily develops a discontinuity, signaling a limitation of Israel-Stewart theory in the ultra-relativistic regime.

\subsection{Asymptotic Stability Analysis}
\label{Linear_1}

The asymptotic equilibrium states define fixed points of the system at $x\rightarrow \pm \infty $, while smooth shock profiles correspond to trajectories connecting them. To determine whether such trajectories can exist, we perform a linear stability analysis \cite{lax_1957,lax_1971,lax_1973} around the asymptotic states, 
\begin{equation*}
\left\{ \varepsilon _{i/f},v_{i/f},\pi =0^{+}\right\} .
\end{equation*}
The conservation laws reduce to the Rankine-Hugoniot relations and therefore provide constants of motion. The asymptotic dynamics is thus governed entirely by the dissipative sector, whose linearized equations take
the form,
\begin{equation}
\left[ \left( v^{2}-\frac{1}{3}\right) \tau _{\pi }-\tau _{\eta }\right]
\partial _{\omega }\delta \hat{\pi}+\frac{1}{\gamma v}\left( v^{2}-\frac{1}{3%
}\right) \delta \hat{\pi}=0,
\end{equation}%
where the velocity gradients were removed using the linearized momentum conservation law, written in the following form,
\begin{equation}
\left( v^{2}-\frac{1}{3}\right) \gamma ^{2}\partial _{\omega }\left( \frac{\delta v}{v}\right) +\partial _{\omega }\delta \hat{\pi}=0.
\end{equation}
For the sake of convenience, we defined $\tau_{\eta }=\eta/
\varepsilon$ and used the notation $\delta \hat{A}=\delta A/\left( \varepsilon +p\right) $.

We seek normal-mode solutions of the form $\sim \exp \left( \lambda \omega\right)$, leading to the following algebraic equation for $\lambda$,
\begin{equation}
\left( v^{2}-\frac{1}{3}\right) \left( \tau _{\pi }+\frac{1}{\lambda \gamma v}\right) -\tau _{\eta }=0,
\end{equation}
with solution
\begin{equation}
\frac{1}{\lambda }=-\gamma v\tau _{\pi }\frac{v^{2}-v_{\mathrm{L}}^{2}}{v^{2}-\frac{1}{3}},
\end{equation}
where we recall that $v_{ \mathrm{L}}$ is the maximum propagation speed of the linearized theory. We note that a similar result was derived in Eq.~(23) of Ref.~\cite{Calzetta_2022}.

This analysis shows that smooth stationary shock solutions may exist as long as $\left\vert v\right\vert <\left\vert v_{\mathrm{L}}\right\vert$. Admissible perturbations must decay toward their corresponding asymptotic equilibrium states. For the upstream state, $v=v_{i}$, we have that $
v_{i}^{2}<1/3$ and $\lambda $ is positive, leading to an exponentially decaying mode in the asymptotic direction $\omega \rightarrow -\infty$. Thus, it satisfies the required decay condition and remains bounded in the asymptotic region. In contrast, at the downstream state, $v=v_{f}$, we have that $v_{f}^{2}>1/3$ and $\lambda $ is negative, corresponding to an exponentially decaying mode in the asymptotic
direction $\omega \rightarrow \infty$. This perturbation also remains bounded in the asymptotic region. The perturbations are therefore asymptotically stable in the directions pointing away from the shock, while growing toward the interior of the profile. This structure is essential for the existence of stationary shock solutions, as it permits trajectories to connect the upstream and downstream equilibrium states. 

This analysis shows that the asymptotic mode structure remains compatible with smooth stationary shock solutions as long as the flow remains in the regular regime. We note that if $\left\vert v\right\vert >\left\vert v_{\mathrm{L}}\right\vert $, the asymptotic mode of the downstream equilibrium acquires the wrong sign, so that no trajectory can approach the downstream state while satisfying the required asymptotic boundary conditions. This is fully consistent with the nonlinear analysis above, where the stationary equations lose regularity precisely at $\left\vert v\right\vert =\left\vert v_{\mathrm{NL}}\right\vert $. Beyond this point, the shock profile can only be completed through a weak solution containing an embedded discontinuity.

\section{Shock solutions}
\label{SecIV}

In this section we investigate the nonlinear shock solutions predicted by the previous analysis. Our primary goal is to determine how the stationary shock structure changes as the asymptotic shock velocity approaches the maximum characteristic velocity of the theory and to assess the consequences of the loss of regularity identified in Sec.~\ref{Sec:ShockWave}. In particular, we examine whether the breakdown of smooth stationary solutions is associated with the formation of embedded discontinuities within the shock profile. We first construct stationary solutions semi-analytically by integrating the nonlinear stationary equations. This allows us to visualize the breakdown mechanism derived in Sec.~\ref{Sec:breakdown} and to identify the location at which the stationary profile becomes singular. We then study the relativistic Riemann problem to determine whether the resulting composite shock structures emerge dynamically from a full time-dependent evolution. For this purpose, we solve the complete hydrodynamic equations numerically using \emph{MUSIC} \cite
{Music1_Schenke_2010,Music2_Schenke_2011,Music3_Paquet_2016}  and verify whether these discontinuities are genuine dynamical solutions of Israel-Stewart theory.

\subsection{Semi-analytical solutions}

In this section we determine the stationary shock profiles of Israel-Stewart theory numerically by integrating the stationary equations
using a second-order Runge-Kutta algorithm. We consider the following boundary conditions: the constants of motion are $C_{1}=-189.451$ fm$^{-4}$ and $C_{2}=165$ fm, leading to the asymptotic velocities, $v_f=-0.642$ and $v_i=-0.519$, and asymptotic energy densities, $\varepsilon_f=130$ fm$^{-4}$ and $\varepsilon_i=200$ fm$^{-4}$. Since the asymptotic equilibrium states are fixed points of the system, a small perturbation is required to initiate the numerical integration away from equilibrium. We thus displace the downstream velocity by $v=v_i - 10^{-4}$ at $\omega _{0}=-3$ fm and recalculate the energy density using the exact relation,
\begin{equation}
   \varepsilon = \frac{C_1}{v} - C_2 .
\label{eq:Energia_em_funcao_velocidade}
\end{equation}
The shear-stress tensor component is then calculated using Eq.~\eqref{EquacaoChoqueGrande}. According to the linear analyses performed in the previous section, this perturbation of the asymptotic state should lead to a shock solution once evolved towards negative values of $\omega$.

Since we are dealing with a conformal fluid, the shear viscosity is taken to be proportional to the entropy density, $s$, 
\begin{equation}
\eta =a_\pi s,
\end{equation}%
with $a_\pi$ being a constant that determines the intensity of dissipation. The
relaxation time is given by the following parametrization, 
\begin{equation}
\tau _{\pi }=b_\pi \frac{\eta }{\varepsilon +p},  \label{taupi}
\end{equation}%
with $b_\pi$ being a dimensionless constant. This parametrization is inspired by kinetic theory calculations \cite{Denicol_PhysRevD.85.114047, Denicol:2014vaa}. 

\begin{figure}[]
\begin{subfigure}[b]{\linewidth}
\centering
\includegraphics[width=7cm]{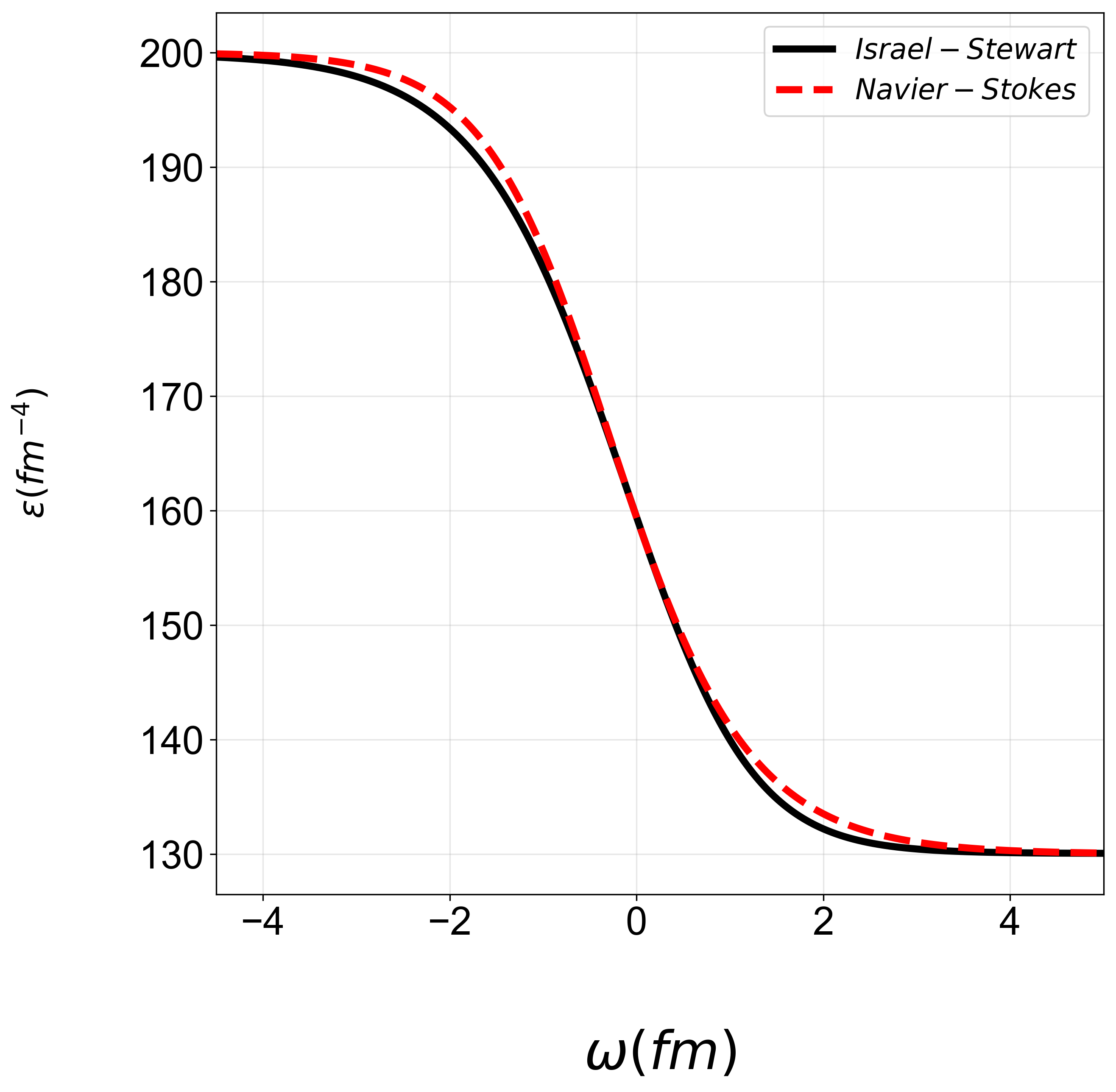}
\end{subfigure}
\begin{subfigure}[b]{\linewidth}
\centering
\includegraphics[width=7cm]{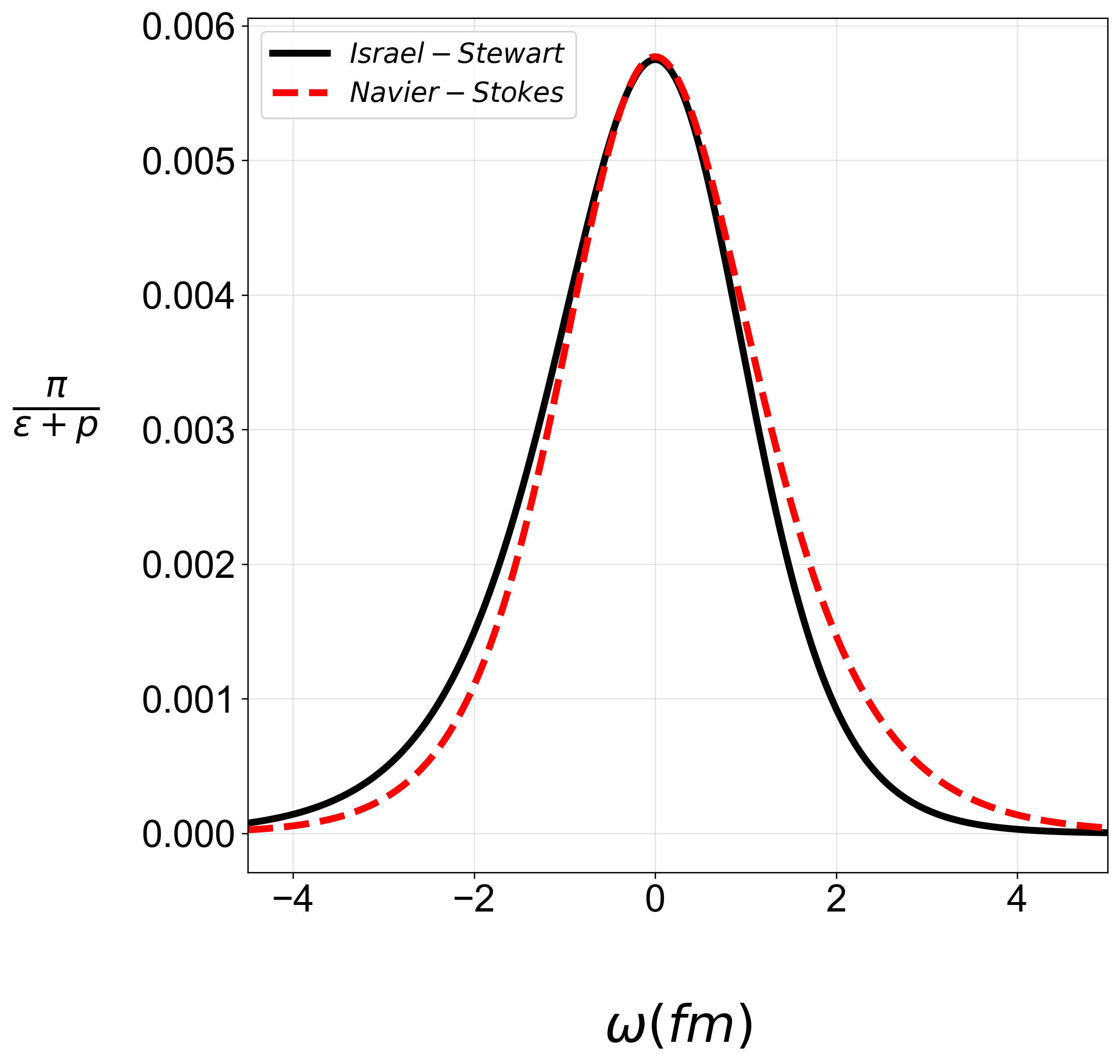}
\end{subfigure}
\caption{Comparison between stationary shock solutions of Israel-Stewart and
Navier-Stokes theory, with $\eta/s = 0.1$ and $b_\pi=5$.}
\label{Comp_Shockwave_IS_vs_NS_a=0_05_shifted}
\end{figure}

We start by comparing the stationary shock solutions of Navier-Stokes and Israel-Stewart theories in a regime where smooth profiles are expected to exist. For the sake of comparison, we shift all solutions in such a way  that the maximum of the shear-stress occurs at $\omega=0$.  The shear viscosity is fixed to $\eta/s = 0.1$, while the relaxation time of Israel-Stewart theory is determined by $b_\pi = 5$. The resulting stationary solutions are shown in Fig.~\ref{Comp_Shockwave_IS_vs_NS_a=0_05_shifted}. Both theories produce smooth viscous shock profiles with a very similar overall structure. However, the presence of a finite relaxation time in Israel-Stewart theory delays the response of the shear-stress tensor to local gradients, leading to a broader shock profile when compared to Navier-Stokes theory. Despite this quantitative difference, the two solutions remain qualitatively similar, indicating that for these parameters the shock velocity remains well within the characteristic propagation regime of the theory and no loss of regularity occurs.

\begin{figure}[]
\begin{minipage}[]{\linewidth}
\centering
\includegraphics[width=7cm]{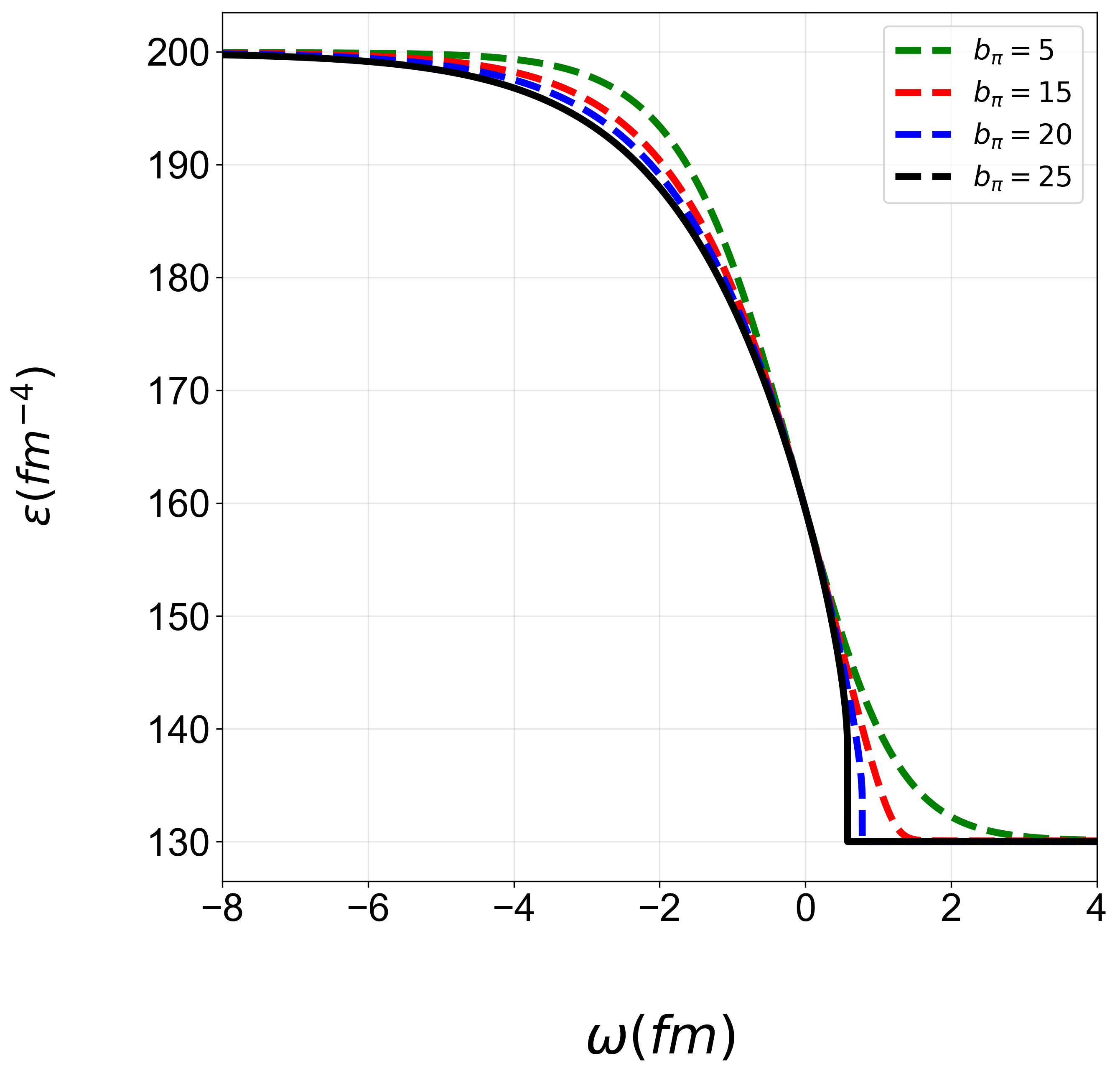}
\end{minipage} 
\begin{minipage}[]{\linewidth}
\centering
\includegraphics[width=7cm]{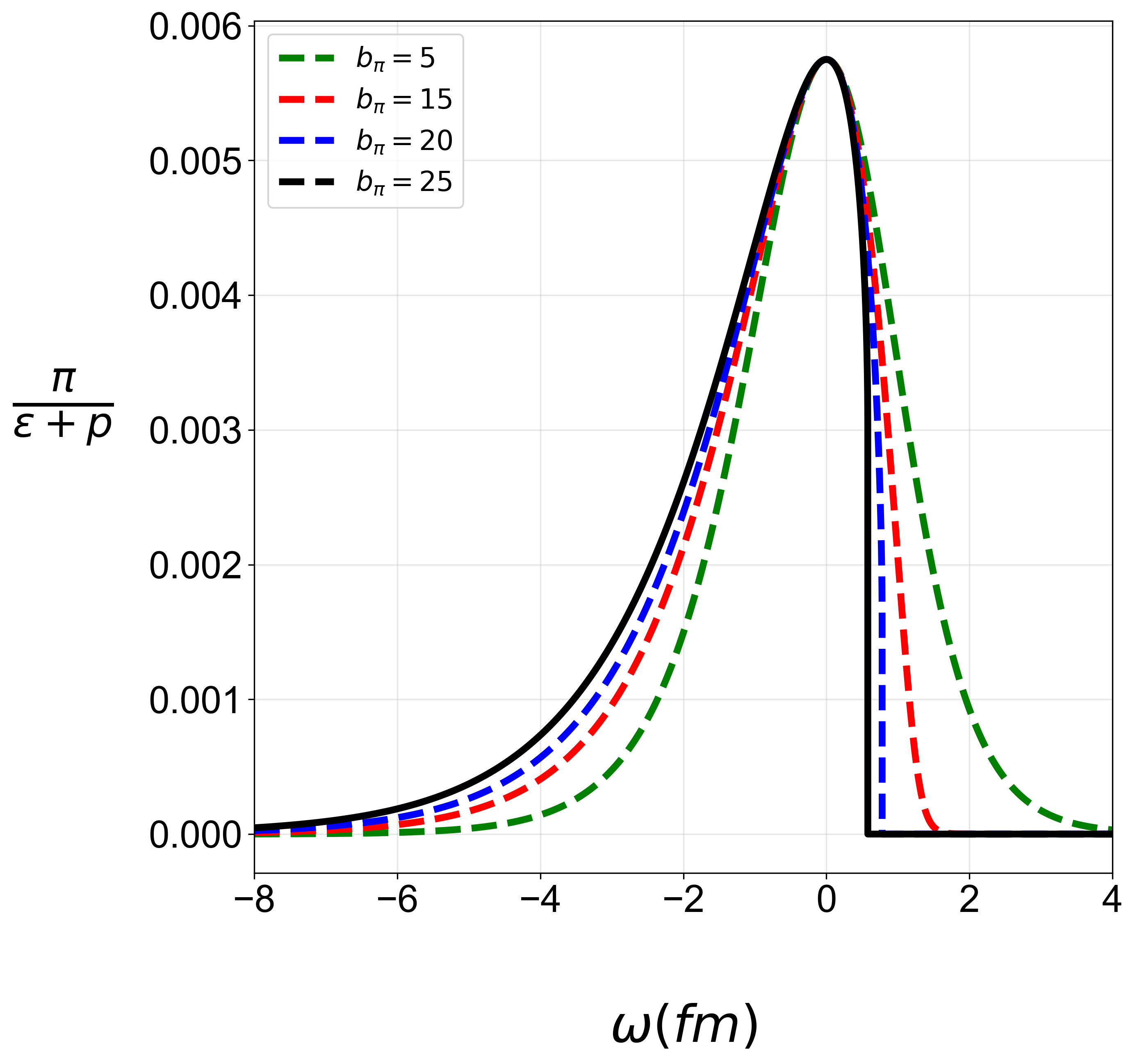}
\end{minipage}
\caption{Stationary shock solutions of Israel-Stewart theory with $\protect%
\eta / s = 0.1$ and different values of relaxation time ($b$).}
\label{Exeplos ShockWave}
\end{figure}

We now test the central prediction of the previous section: smooth stationary shock profiles should cease to exist once the shock velocity exceeds the characteristic propagation speed supported by the theory. Rather than increasing the shock strength directly, we probe this transition by varying the parameter $b_\pi$, which modifies the relaxation time and therefore changes the characteristic propagation velocity of Israel-Stewart theory while keeping the asymptotic states fixed. The resulting stationary profiles of the energy density and $\pi/p$ as a function of $\omega$ are shown in Fig.~\ref{Exeplos ShockWave}. For $b_\pi=5$ and $b_\pi=15$, the characteristic propagation velocity remains larger than the shock velocity, $|v_{ \mathrm{NL}}|>|v_{\text{shock}}|$,
and the stationary solutions remain smooth throughout the entire profile. In contrast, for $b_\pi=20$ and $b_\pi=25$, the shock velocity exceeds the characteristic propagation speed,
$|v_{ \mathrm{NL}}|<|v_{\text{shock}}|$,
and the stationary equations lose regularity at a finite value of $\omega$. The smooth branch of the solution then terminates and must be completed by a discontinuity satisfying the Rankine-Hugoniot conditions.
This transition is highly nontrivial. Naively, one might expect the divergence of the velocity gradient to be accompanied by correspondingly large dissipative corrections. However, the solutions show that the dissipative stresses remain finite and that the nonlinear characteristic velocity differs only modestly from its linearized value. The onset of singular behavior is therefore not driven by an uncontrolled growth of nonequilibrium corrections. Instead, it occurs due to the inability of the causal hydrodynamic theory to propagate information rapidly enough to continuously resolve the stationary shock structure. In this sense, the breakdown identified here differs qualitatively from the conventional picture in which hydrodynamics fails due to large dissipative corrections \cite{Dusling_2008}.
\begin{figure}[]
\begin{minipage}{\linewidth}
\includegraphics[width=7cm]{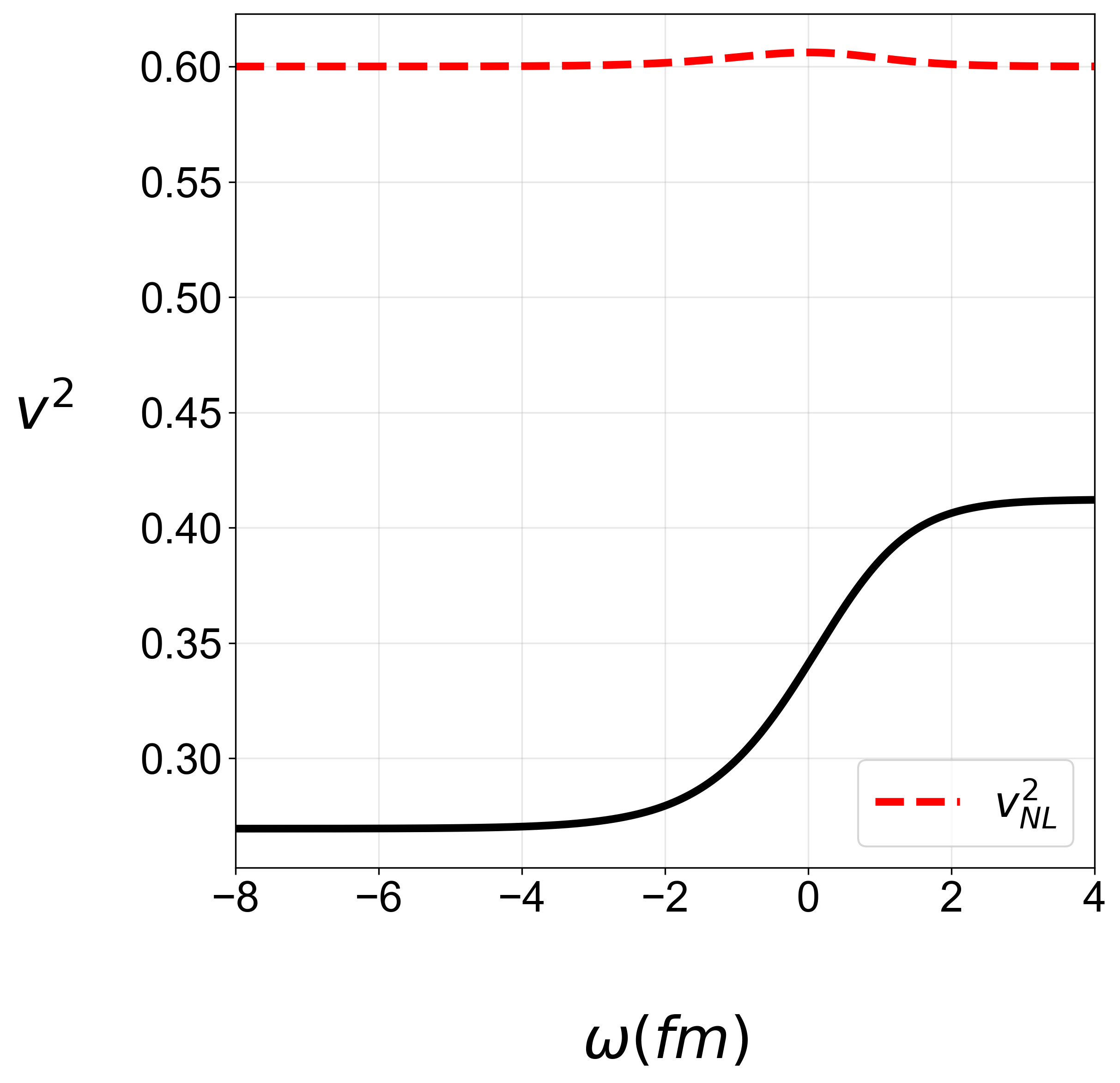}
\subcaption{$b_\pi = 5$}
\end{minipage}%
\vspace{0.1cm}
\begin{minipage}{\linewidth}
\includegraphics[width=7cm]{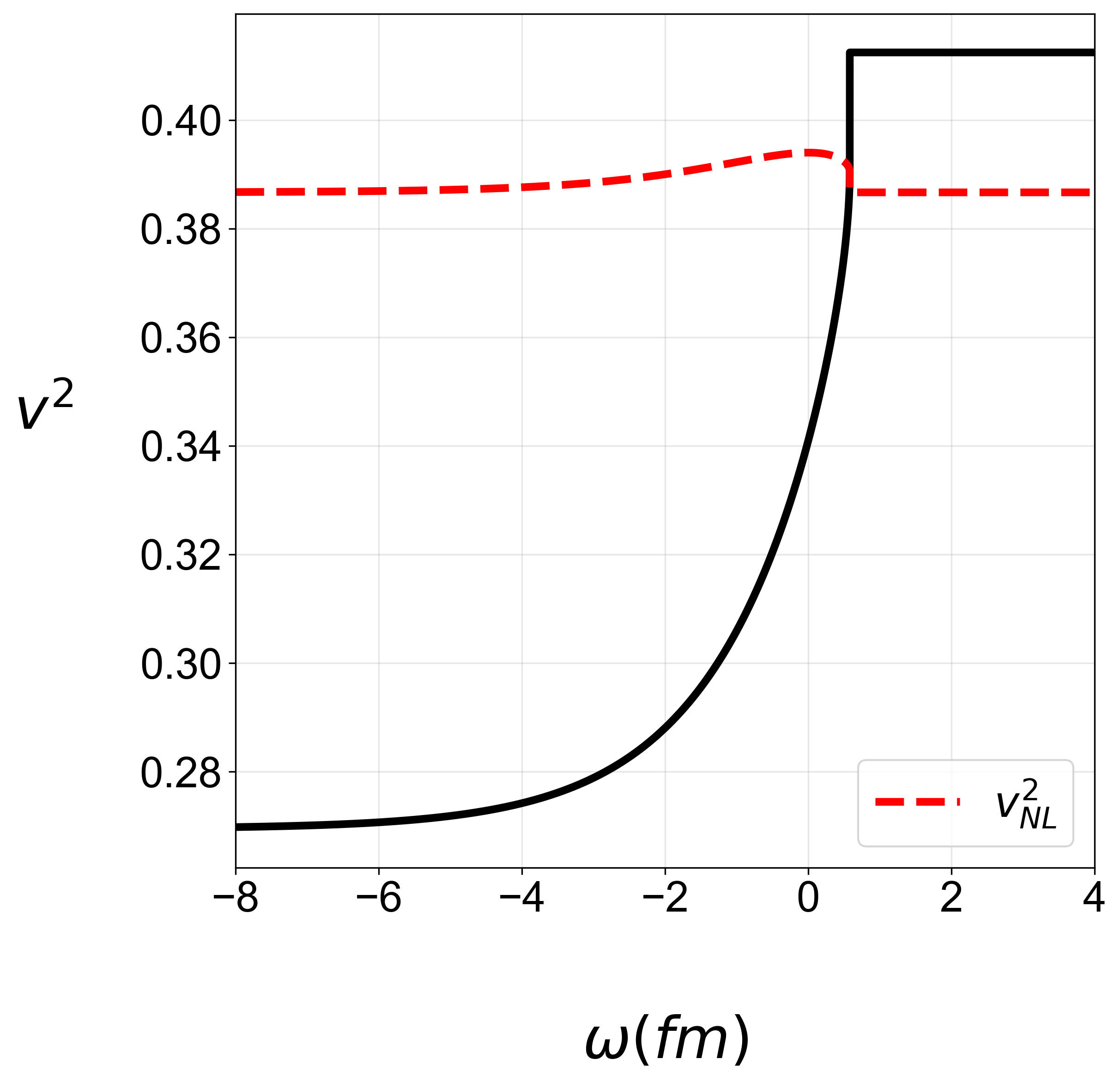}
\subcaption{$b_\pi = 25$}
\end{minipage}
\caption{Comparison of the velocities and the maximum propagation speed for different stationary shock solutions
with $\eta /s=0.1$}
\label{fig:Comp_shock_vs_velocidades_min_e_max}
\end{figure}

Figure \ref{fig:Comp_shock_vs_velocidades_min_e_max} provides a direct visualization of the breakdown mechanism derived in Sec.~\ref{Sec:breakdown} by displaying $v^2$ and $v_{\textrm{NL}}^2$ as a function of $\omega$. For $b_\pi=5$, the characteristic velocity remains larger than the flow velocity throughout the profile and the solution stays smooth. In contrast, for $b_\pi=25$, the flow velocity reaches the nonlinear characteristic velocity at a finite value of $\omega$. The loss of regularity occurs precisely at this intersection, 
where the smooth profile terminates and must be continued by an embedded discontinuity. This figure also shows that the nonlinear corrections to the characteristic velocity remain relatively small across the profile, with $v^2_{\textrm{NL}}$ remaining approximately constant as a function of $\omega$. The breakdown mechanism is thus predominantly controlled by the linear propagation structure of the theory, while nonlinear dissipative corrections provide only modest quantitative modifications. 

\subsection{Riemann problem}
\label{Numerical_Simulations}
In this section we verify whether the stationary shock solutions discussed in the previous section can be formed dynamically via the Riemann problem. 
For this purpose, we solve Israel-Stewart theory numerically using the hydrodynamic simulation \emph{MUSIC} \cite
{Music1_Schenke_2010,Music2_Schenke_2011,Music3_Paquet_2016}. First, we introduce the initial condition that will lead to a
stationary shock solution, known as the Riemann problem \cite%
{Riemann_loraclavijo2013exact}. At the initial time, the fluid is
separated into two regions, with an interface at $x=0$, each with a different energy density, $\varepsilon_1=331.48$ fm$^4$ and 
$\varepsilon_2=131.52$ fm$^4$, vanishing velocity and vanishing shear-stress tensor. The interface is then removed, and as the system attempts to equilibrate, a shock wave emerges.
We adopt
a grid size of $dx = 0.01$ fm and a time step of $dt=0.01$ fm \footnote{\emph{MUSIC} solves the hydrodynamic equations using hyperbolic coordinates and we effectively convert it to cartesian coordinates by setting the initial hyperbolic time to $\tau_0 = 10^6$ fm. In this limit, $\tau \rightarrow t$.
}. 

\begin{figure}[]
\begin{minipage}[]{\linewidth}
\centering
\includegraphics[width=7cm]{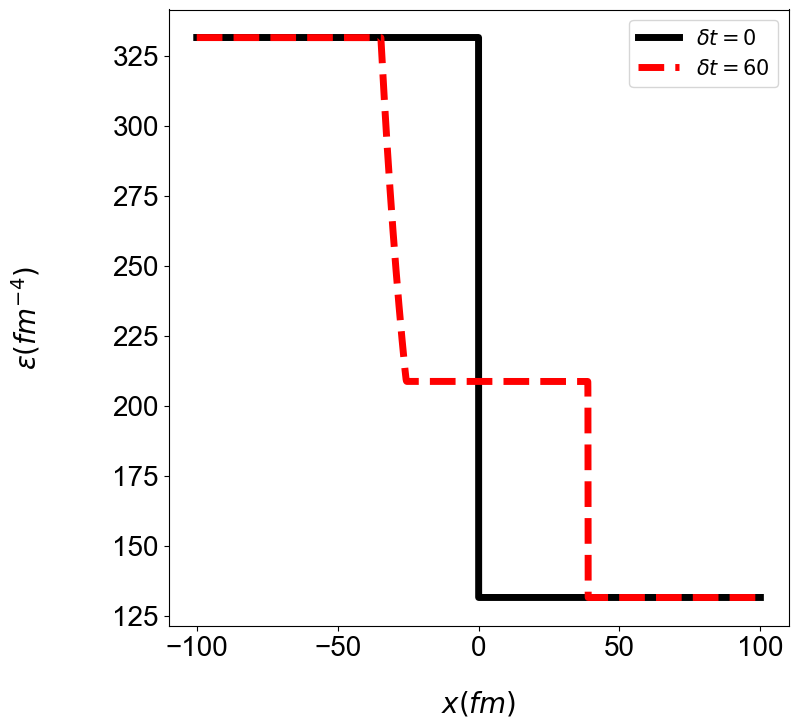}
\end{minipage}
\begin{minipage}[]
{\linewidth}
\centering
\includegraphics[width=7cm]{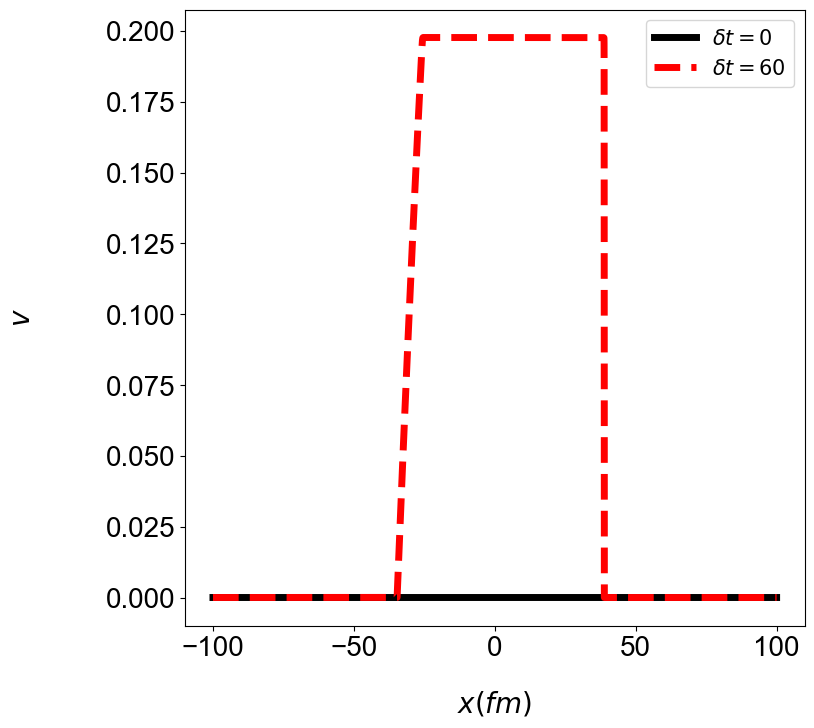}
\end{minipage}
\caption{Time evolution for Riemann initial condition.}
\label{Riemann}
\end{figure}

For the sake of completeness, we first consider the corresponding ideal-fluid Riemann problem, shown in Fig.~\ref{Riemann}. The solution exhibits the familiar self-similar structure consisting of a rarefaction wave propagating to the left, a uniform intermediate state, and a shock wave propagating to the right. The shock connects the intermediate state to the upstream fluid and is the structure of primary interest for the present work. It is important to note that, unlike the stationary solutions discussed in the previous section, the Riemann solution is displayed in the laboratory frame rather than in the rest frame of the shock. Consequently, the shock appears as a propagating discontinuity and is accompanied by the additional wave structures generated by the initial-value problem. The stationary profiles derived previously correspond only to the internal structure of the shock itself once transformed to the shock rest frame. For the ideal fluid, the shock profile is simply the discontinuity predicted by the Rankine-Hugoniot conditions, slightly broadened by numerical viscosity. We emphasize that the strength of the shock is determined by the intermediate state generated by the Riemann problem and is therefore not directly equal to the initial density ratio $\varepsilon_1/\varepsilon_2$ imposed at $t=0$. We see from the figure that the emerging shock wave has $\varepsilon_i=208.68$ fm$^4$ and
$\varepsilon_f = \varepsilon_2$. 

We now consider dissipative fluids. Our primary goal is to determine whether the composite shock structures predicted by the stationary analysis can be realized dynamically in Israel-Stewart theory. To this end, we solve the time-dependent equations for a fixed shear viscosity $\eta/s=0.1$ and two values of the relaxation parameter, $b_\pi=5$ and $b_\pi=20$. The initial condition is identical to the one employed in the ideal-fluid Riemann problem. In order to facilitate comparison with the stationary solutions constructed in the previous section, we restrict our analysis to the shock-wave component resulting from the Riemann problem and transform the profiles to the local rest frame of the shock. 

\begin{figure}[]
\begin{subfigure}[b]{\linewidth}
\centering
\includegraphics[width=7cm]{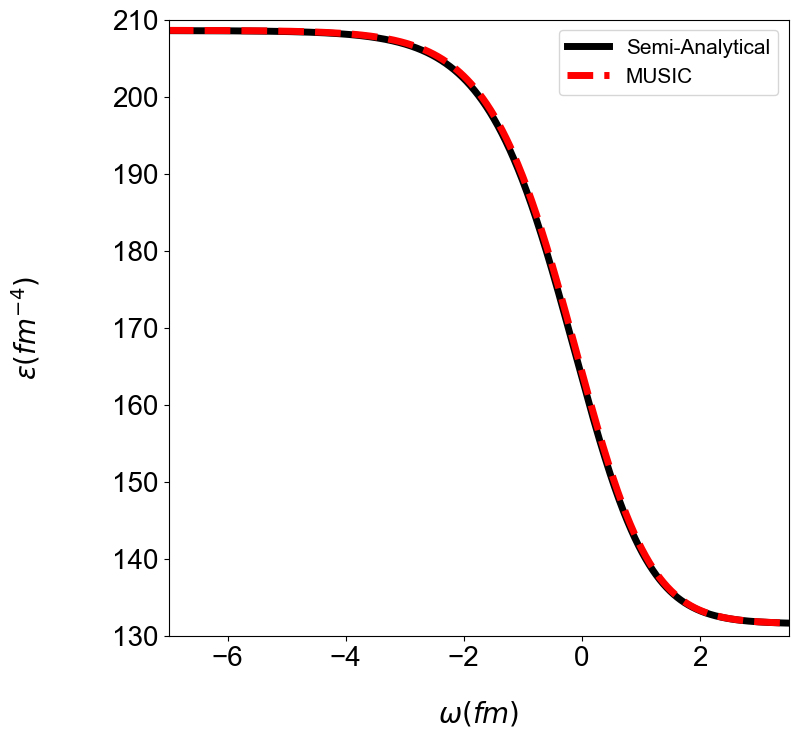}
\end{subfigure}\
\begin{subfigure}[b]{\linewidth}
\centering
\includegraphics[width=7cm]{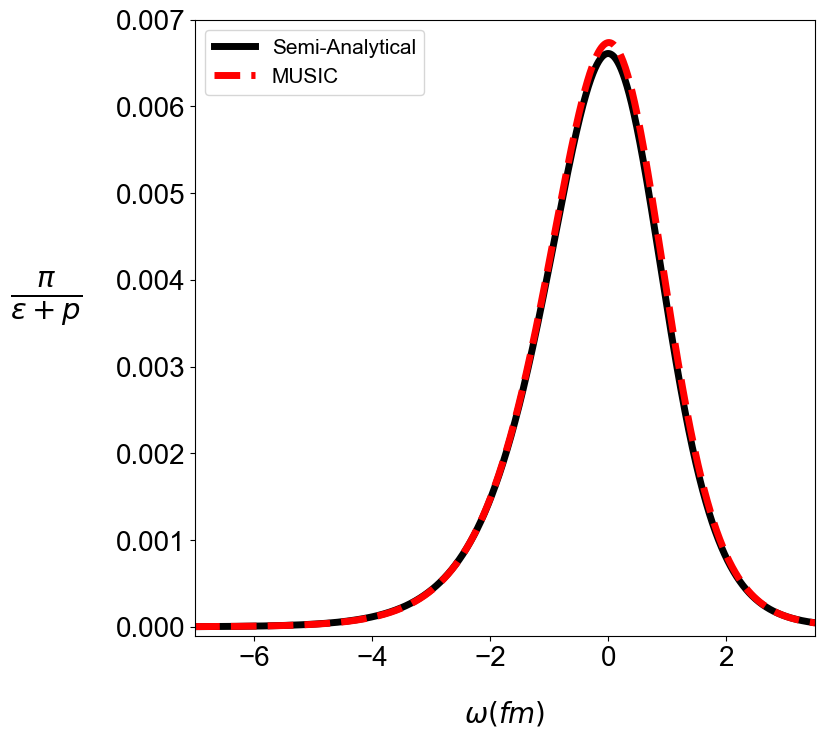}
\end{subfigure}
\caption{Comparison between continuous shock solution, one using MUSIC and
another obtained semi-analytically, with $\eta/s=0.1$, $b_\pi=5$.}
\label{Shock_b}
\end{figure}

\begin{figure}[]
\begin{minipage}{\linewidth}
\centering
\includegraphics[width=7cm]{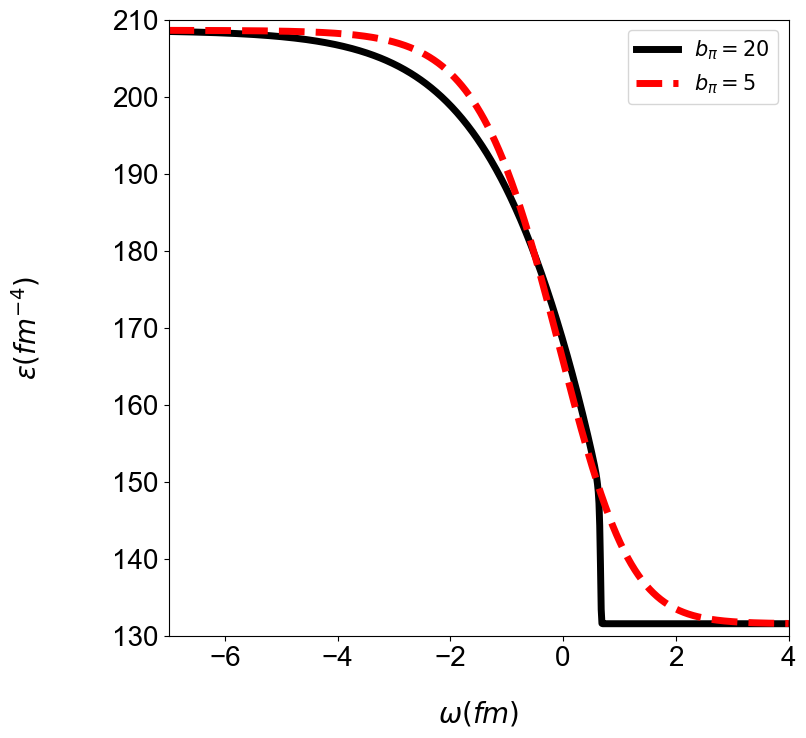}
\end{minipage}
\begin{minipage}{\linewidth}
\centering
\includegraphics[width=7cm]
{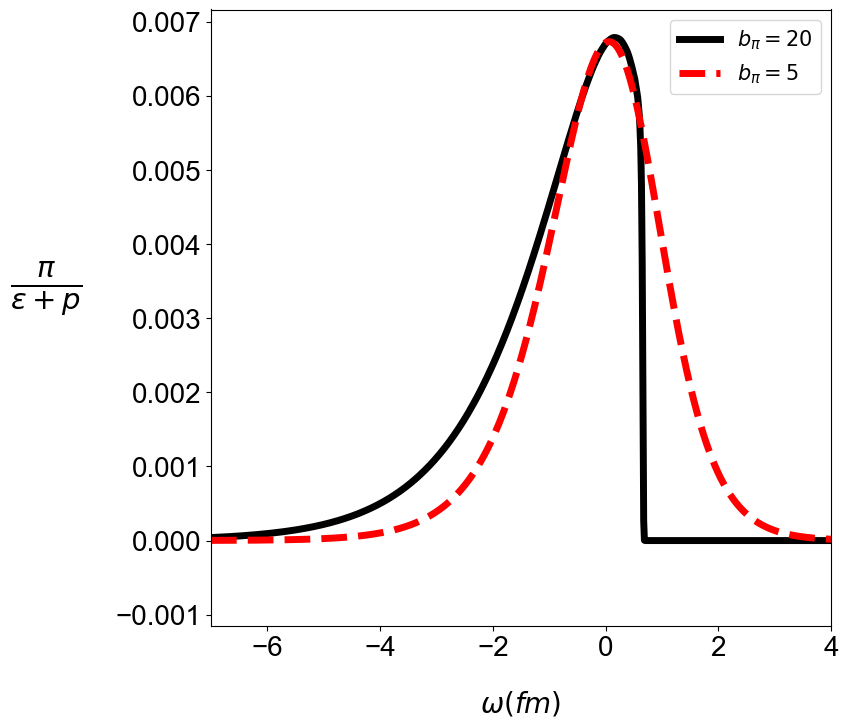}
\end{minipage}
\caption{Comparison of how a stationary shock wave propagates in different
fluids, each with different relaxation time, made with \emph{MUSIC}, with $%
\eta/s=0.1$.}
\label{Shock_b2}
\end{figure}

\begin{figure}[]
\centering
\begin{subfigure}{\linewidth}
\includegraphics[width=7cm]{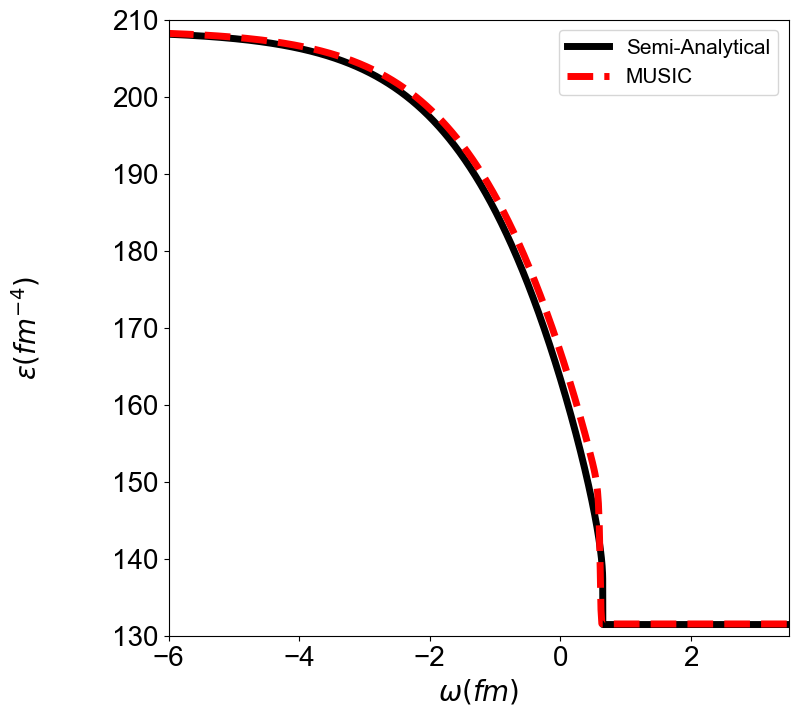}
\end{subfigure}
\begin{subfigure}{\linewidth}
\includegraphics[width=7cm]{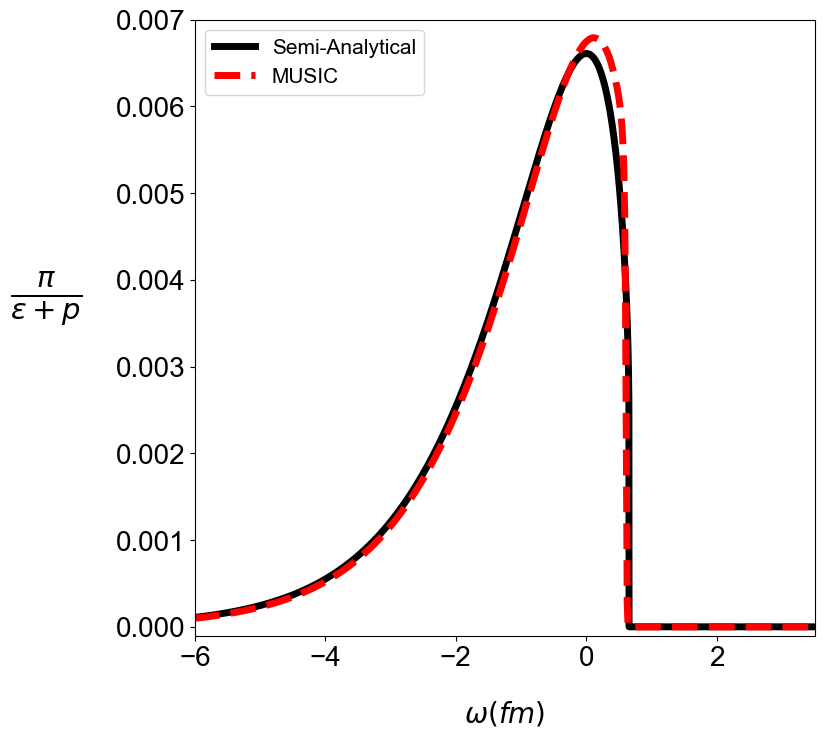}
\end{subfigure}
\caption{Comparison between solutions of MUSIC and semi-analytical results with $\eta/s=0.1$, $b_\pi=20$.}
\label{Music_Singular_Comparison}
\end{figure}

Figure \ref{Shock_b} shows the shock profile obtained for $b_\pi=5$, where the stationary analysis predicts the existence of a smooth solution. We indeed find that the shock remains smooth and that the profile extracted from the numerical simulation is in excellent agreement with the semi-analytical stationary solution constructed in the previous section. We now turn to the case of $b_\pi=20$, for which the stationary analysis predicts the breakdown of smooth shock profiles because the shock velocity exceeds the maximum propagation speed supported by the theory. The corresponding results are shown in Fig.~\ref{Shock_b2}, where solutions obtained using \emph{MUSIC} for $b_\pi=5$ and $b_\pi=20$ are displayed. In agreement with our stationary wave analyses, the numerical solution obtained for $b_\pi=20$ develops an additional singular structure within the shock profile. Moreover, the location and overall shape of this structure are in good agreement with the semi-analytical composite shock solution, as shown in Fig.~\ref{Music_Singular_Comparison}.

This agreement demonstrates that the embedded discontinuity emerges naturally from the full time-dependent evolution of the Israel-Stewart equations. Whether such structures can arise in physical systems remains an open question and will likely require microscopic studies, for example through kinetic theory simulations, in order to establish the range of validity of the hydrodynamic description in this regime. An initial study of this problem was performed in the linear regime in Ref.~\cite{Calzetta_2022}, solving the Boltzmann equation in the relaxation time approximation, and indicated that such discontinuities do not emerge in kinetic theory.

Recent work by Disconzi, Hoang and Radosz \cite{Disconzi_2023} established that smooth solutions of Israel-Stewart theory do not, in general, remain globally regular. More precisely, they proved the existence of smooth and physically admissible initial data for which the solution cannot be continued as a smooth admissible solution for all times. Their theorem, however, does not identify the mechanism responsible for the breakdown. The analysis only implies that either derivatives become singular or the solution reaches the boundary of the admissible state space, leaving open the question of how regularity is ultimately lost. 

The present work identified a concrete nonlinear mechanism through which such a loss of regularity can occur in the context of shock waves. We have shown that smooth stationary shock profiles cease to exist when the shock velocity reaches a characteristic propagation velocity of the nonlinear theory. At this point the stationary equations become singular because the shock trajectory intersects a characteristic hypersurface, preventing a smooth continuation of the profile. While our analysis is restricted to stationary solutions and therefore does not establish the mechanism responsible for finite-time breakdown in the general initial-value problem, it suggests that characteristic-speed saturation may provide a natural route through which regularity is lost in strongly nonlinear relativistic flows.

Finally, we note that the Riemann problem was also investigated in the context of Israel-Stewart theory in Ref.~\cite{Bemfica_2025}. That work demonstrated the existence of smooth initial data that develop finite-time gradient singularities and provided numerical evidence that these singularities evolve into physical shock waves satisfying the Rankine-Hugoniot conditions. The focus there was on the fully dynamical, early-time regime in which nonlinear wave steepening dominates viscous relaxation, rather than on the late-time stationary shock structures considered here. Nevertheless, it provides another example of the emergence of shock-like discontinuities in Israel-Stewart theory. Together with the present results, it suggests that finite-speed propagation may play an important role in the loss of regularity of strongly nonlinear solutions.

\section{Third-order hydrodynamics}
\label{SecV}

The breakdown mechanism identified above suggests that the obstruction to
smooth shock solutions is not solely associated with the magnitude of
gradients, but also with the ability of the theory to transport information
across the shock profile: smooth stationary
solutions cease to exist exactly when the shock velocity reaches the maximum
propagation speed supported by the hydrodynamic equations. This observation suggests a possible route toward regularization. In derivations of relativistic hydrodynamics from kinetic theory using the
method of moments \cite{Denicol_PhysRevD.85.114047,Denicol2021-oe}, the dissipative currents are not isolated dynamical
variables. Instead, the shear-stress tensor is
coupled to an infinite hierarchy of higher-order moments of the single-particle distribution
function. Israel-Stewart theory emerges only after truncating this
hierarchy and expressing the higher moments in terms of lower-order
variables through a systematic power-counting scheme \cite{Denicol_PhysRevD.85.114047}.

Under ordinary hydrodynamic conditions, this truncation is justified because
the neglected moments contribute only at higher orders in the Knudsen and
inverse Reynolds expansions. In this regime, gradients are sufficiently
small and the additional moments are expected to relax rapidly without
significantly modifying the dynamics of the long-wavelength modes. The shock
solutions studied here indicate that this picture becomes more subtle in
strongly relativistic flows. The existence of smooth stationary solutions is
controlled not only by the magnitude of gradients, but also by the
characteristic propagation speeds of the truncated theory. Even when the dissipative corrections remain finite, the hydrodynamic description can break down if the theory cannot propagate information sufficiently rapidly
to resolve the shock profile continuously. From this perspective, the neglected higher-order moments may become
dynamically relevant not because they dominate the local constitutive
relations, but because they modify the hyperbolic structure of the equations
and enlarge the range of accessible propagation speeds.  This interpretation is supported by the analysis of Ref.~\cite{Calzetta_2022}, where a divergence-type theory incorporating the third moment of the Boltzmann distribution was shown to possess a larger characteristic velocity than the corresponding second-moment (Israel-Stewart-like) description, thereby extending the range of shock velocities for which smooth stationary solutions exist.

To investigate this possibility within the framework of kinetic theory, we consider the simplest extension of Israel-Stewart theory obtained by promoting the lowest neglected kinetic moment to an independent dynamical variable \cite{debrito2023thirdorder}. Within this formalism, the lowest-order degree of freedom neglected in the 14-moment approximation is a rank-three irreducible
tensor, which we denote by $\Omega^{\mu\nu\rho}$, a symmetric, traceless rank-three tensor orthogonal to the fluid four-velocity. Following the results of Ref.~\cite{debrito2023thirdorder},
we consider a new theory that couples the evolution of the shear-stress tensor to this additional moment, allowing gradients of $\pi^{\mu\nu}$ to influence the dynamics through an extra relaxation channel. The equation of motion for the shear-stress tensor is extended to,
\begin{eqnarray}
\tau _{\pi }\Delta_{\alpha }^{\mu}
\Delta_{\beta}^{\nu}
u^{\lambda }\partial _{\lambda }\pi
^{\alpha \beta }+\pi ^{\mu \nu }
=2\eta\sigma ^{\mu \nu }
-\frac{4}{3}\tau _{\pi }\pi ^{\mu \nu }\partial _{\lambda }u^{\lambda }
\notag \\
-\tau _{\pi }\Delta_{\alpha }^{\mu}
\Delta_{\beta}^{\nu}\nabla _{\lambda }\Omega
^{\lambda \alpha \beta },    
\end{eqnarray}
 The new moment obeys a relaxation-type equation in which gradients of the shear-stress tensor act as source terms,
\begin{equation}
\tau _{\Omega }
\Delta_{\alpha }^{\mu}
\Delta_{\beta}^{\nu}
\Delta_{\chi }^{\rho}
u^{\lambda }\partial _{\lambda }\Omega ^{\alpha \beta \chi }
+\Omega ^{\mu \nu \rho }
=
\frac{3}{7}\eta _{\Omega }
\Delta _{\alpha \beta \chi }^{\mu \nu \rho }
\nabla ^{\alpha }\pi ^{\beta \chi },
\end{equation}
 where $\Delta ^\lambda_\nu = \delta ^\lambda_\nu - u^\lambda_\nu $ and $\Delta _{\alpha \beta \chi }^{\mu \nu \rho }$ is a triple, symmetric and traceless projection operator \cite{debrito2023thirdorder}. We note that the complete theory contains several additional nonlinear source terms that contribute
to the evolution of $\Omega^{\mu\nu\rho}$ \cite{debrito2023thirdorder}. Here, we retain only the coupling
that directly modifies the characteristic propagation speeds of the hydrodynamic equations. This simplified model allows us to isolate the impact of additional dynamical degrees of freedom on the causal structure of
the theory and, consequently, on the existence of smooth stationary shock
solutions.

In the one-dimensional limit, the equations then become,%
\begin{eqnarray}
u\partial _{\omega }\pi +\frac{\pi }{\tau _{\pi }} &=&-\frac{4}{3}\left( 
\frac{\eta }{\tau _{\pi }}+\pi \right) \partial _{\omega }u-\gamma \partial
_{\omega }\Omega , \\
u\partial _{\omega }\Omega +\frac{\Omega }{\tau _{\Omega }} &=&-\frac{3\eta
_{\Omega }}{5\tau _{\Omega }}\gamma \partial _{\omega }\pi ,
\end{eqnarray}%
where $\Omega \equiv \Omega^{xxx}/\gamma^{3}$. In Ref.~\cite{Brito_2020},
the following expressions for the transport coefficients were derived 
\begin{eqnarray}
\tau _{\Omega } &=&\tau _{\pi }, \\
\eta _{\Omega } &=&\frac{18}{49}\tau _{\pi }.
\end{eqnarray}%
In the numerical examples discussed below, we shall use these kinetic-theory estimates for the transport coefficients.

\subsection{Stationary wave solution and regularity}

We now use the conservation laws and remove the energy density and velocity
field gradients from the equations of motion for the dissipative currents.
The dissipative equations can be reduced to, 
\begin{eqnarray}
\left( u^{2}-\frac{3\eta _{\Omega }}{5\tau _{\pi }}\gamma ^{2}\right)
\partial _{x}\pi +\frac{u\pi }{\tau _{\pi }} &=&-\frac{4}{3}
\eta_\pi u\partial _{x}u+\frac{\gamma \Omega }{\tau
_{\pi }}, \\
u\partial _{x}\Omega +\frac{\Omega }{\tau _{\pi }} &=&-\frac{3\eta _{\Omega }%
}{5\tau _{\pi }}\gamma \partial _{x}\pi, 
\end{eqnarray}%
where we defined $\eta_\pi = (\eta/\tau_\pi + \pi)$ and then reduce it to one single equation of the following form, 
\begin{equation}
\left( \varepsilon +p+\pi \right) \left( v^{2}-v_{-}^{2}\right) \left(
v^{2}-v_{+}^{2}\right) \gamma ^{3}\frac{\partial _{x}v}{v}=\frac{v\pi
-\Omega }{\tau _{\pi }},
\label{Master_equation_shear_omega}
\end{equation}
where we defined the minimum and maximum characteristic velocity scales,%
\begin{equation}
v_{\pm }^{2}=\frac{v_{\mathrm{NL}}^{2}+\frac{3\eta _{\Omega }}{5\tau _{\pi }}%
\pm \sqrt{\left( v_{\mathrm{NL}}^{2}+\frac{3\eta _{\Omega }}{5\tau _{\pi }}
\right) ^{2}-\frac{4\eta _{\Omega }}{5\tau _{\pi }}}}{2},
\end{equation}
with $v_{\mathrm{NL}}^{2}$ being the maximum propagating velocity of the
theory with only shear. We note that both velocities are real since $v_{%
\mathrm{NL}}^{2}\geq 1/3$. Furthermore, causality dictates that%
\begin{equation}
v_{+}^{2}\leq 1\Longrightarrow v_{\mathrm{NL}}^{2}+\frac{2\eta _{\Omega }}{%
5\tau _{\pi }}\leq 1.
\end{equation}
If the dissipative currents are set to zero, this reduces to the condition
found in Ref.~\cite{debrito2023thirdorder} for longitudinal perturbations. 

We see that the inclusion of the additional dynamical field introduces a
second velocity scale into the theory. This is a qualitatively new feature
absent in Israel-Stewart theory, where only an upper propagation bound
appears. As in the Israel-Stewart case, the velocity scales $v_{\pm }$
emerging from the stationary shock analysis are expected to coincide with
the characteristic propagation velocities obtained from the local causality
analysis of the dynamical equations. 

The stationary equations become singular whenever the coefficient
multiplying the velocity gradient vanishes, i.e., when 
\begin{equation}
\left( \varepsilon +p+\pi \right) \left( v^{2}-v_{-}^{2}\right) \left(
v^{2}-v_{+}^{2}\right) =0.
\end{equation}
The solutions then become singular if the right-hand side does not vanish,
\[
v\pi -\Omega \neq 0.
\]%
In this case, the problem must be reformulated in terms of weak solutions of
the conservation laws, via the Rankine-Hugoniot conditions \cite{Landau}. In
principle, regular continuation across the characteristic hypersurfaces $%
v^{2}=v_{\pm }^{2}$ could still occur if the right-hand side of the equation
simultaneously vanishes, i.e., if $v\pi -\Omega =0$ at the same point.
However, the extended system contains an additional dynamical degree of
freedom and there is presently no general argument implying that such
simultaneous cancellations occur dynamically. On the contrary, this would
require special compatibility conditions on the stationary flow and
therefore should not be expected generically.

\subsection{Conditions for breakdown of smooth solutions}

In the following, we make this discussion more concrete and discuss solutions under specific boundary conditions. We consider stationary solutions initialized within the sector%
\begin{equation}
\left\vert v_{-}\right\vert <\left\vert v\left( \omega \right) \right\vert
<\left\vert v_{+}\right\vert ,
\end{equation}
together with%
\begin{equation}
v\pi -\Omega <0,
\end{equation}
and asymptotic boundary conditions approaching equilibrium with $%
\lim_{\omega \rightarrow \infty }\pi \left( \omega \right) =0^{+}$ and $%
\lim_{\omega \rightarrow \infty }\Omega \left( \omega \right) =0^{+}$.
Within this sector, the stationary equations remain manifestly regular. Crossing its boundaries is dynamically nontrivial because the stationary equations become degenerate there. We thus search for the conditions to find
solutions that remain in the domain $\left\vert v_{-}\right\vert <\left\vert
v\left( \omega \right) \right\vert <\left\vert v_{+}\right\vert $ and $v\pi
-\Omega <0$, ensuring a sufficient, even though not necessary, condition for
regularity.

The addition of a new field and transport coefficients provide another way
to adjust the propagation speeds of the theory and improve the description
of strong shocks. Nevertheless, the presence of two characteristic velocity
scales implies that regularization cannot be achieved simply by increasing
the maximum propagation speed $\left\vert v_{+}\right\vert $ toward the
speed of light through the transport coefficient $\eta _{\Omega }$.
Increasing $\left\vert v_{+}\right\vert $ simultaneously raises the lower
velocity threshold $\left\vert v_{-}\right\vert $, potentially driving the
upstream region of the shock outside the regularity window. The additional
dynamical sector therefore modifies the admissible propagation structure in
a nontrivial way, constraining both the upstream and downstream regions of
the stationary profile simultaneously. Nevertheless, even if the inclusion
of additional moments cannot completely eliminate the loss of regularity, it
can significantly enlarge the regime in which smooth stationary solutions
exist, increasing the domain
of applicability of relativistic hydrodynamics. We shall demonstrate this explicitly in the numerical
solutions presented below.

\subsection{Linear Stability analysis}

To determine whether regular solutions can indeed be constructed, we perform
a linear stability analysis around the asymptotic equilibrium states. As
already noted, the conservation laws reduce to the Rankine-Hugoniot
relations and therefore provide constants of motion. The asymptotic dynamics
is thus governed entirely by the dissipative sector, which now reads, 
\begin{eqnarray}
\frac{v^{2}-v_{\mathrm{L}}^{2}}{v^{2}-\frac{1}{3}}v\partial _{\omega }\delta 
\hat{\pi}+\frac{\delta \hat{\pi}}{\gamma \tau _{\pi }}+\partial _{\omega
}\delta \hat{\Omega} &=&0, \\
v\partial _{\omega }\delta \hat{\Omega}+\frac{\delta \hat{\Omega}}{\gamma
\tau _{\pi }}+\frac{3\eta _{\Omega }}{5\tau _{\pi }}\partial _{\omega
}\delta \hat{\pi} &=&0,
\end{eqnarray}%
where $\delta \hat{A}=\delta A/\left( \varepsilon +p\right) $ and $v_{%
\mathrm{L}}$ is the maximum propagating velocity of the linearized
theory with only shear viscosity.

We seek normal-mode solutions of the form $\sim \exp \left( \lambda \omega
\right) $, leading to the following characteristic equation for $\lambda $,%
\begin{equation}
\left( v^{2}-v_{\mathrm{L}-}^{2}\right) \left( v^{2}-v_{\mathrm{L}%
+}^{2}\right) \left( \gamma \tau _{\pi }\lambda \right) ^{2}+2\Gamma v\gamma
\tau _{\pi }\lambda +v^{2}-\frac{1}{3}=0,
\end{equation}%
where, 
\begin{equation}
    \Gamma = \left( v^{2}-\frac{1}{3}-\frac{2}{3%
}\frac{\tau _{\eta }}{\tau _{\pi }}\right) ,
\end{equation}
and $v_{\mathrm{L}\pm }^{2}$ is the corresponding linearized version of $%
v_{\pm }^{2}$, respectively,
\begin{equation}
v_{\mathrm{L}\pm }^{2}=\frac{v_{\mathrm{L}}^{2}+\frac{3\eta _{\Omega }}{%
5\tau _{\pi }}\pm \sqrt{\left( v_{\mathrm{L}}^{2}+\frac{3\eta _{\Omega }}{%
5\tau _{\pi }}\right) ^{2}-\frac{4\eta _{\Omega }}{5\tau _{\pi }}}}{2}.
\end{equation}
The equation is solved by
\begin{equation}
\gamma \tau _{\pi }\lambda _{\pm }=\frac{-\Gamma v\pm \sqrt{\Gamma^{2}v^{2}-\left(
v^{2}-v_{\mathrm{L}-}^{2}\right) \left( v^{2}-v_{\mathrm{L}+}^{2}\right)
\left( v^{2}-\frac{1}{3}\right) }}{\left( v^{2}-v_{\mathrm{L}-}^{2}\right)
\left( v^{2}-v_{\mathrm{L}+}^{2}\right) }.
\end{equation}
The discriminant can be verified straightforwardly to be positive definite
as long as $\eta _{\Omega }>0$ and thus both eigenvalues are real.

We now assume that the regularity condition is not violated, i.e., $\left(
v^{2}-v_{\mathrm{L}-}^{2}\right) \left( v^{2}-v_{\mathrm{L}+}^{2}\right) <0$%
, and analyse the upstream and downstream perturbations. For the upstream
asymptotic state we have $v=v_{i}$ and $v_{i}^{2}-\frac{1}{3}<0$, leading to
both eigenvalues with a positive sign, describing an exponentially decaying
mode in the asymptotic direction $\omega \rightarrow -\infty $. Thus, they
satisfy the required decay conditions, implying that nearby trajectories
remain bounded in the asymptotic region. In contrast, at the downstream
state, $v=v_{f}$, we have that $v_{f}^{2}>1/3$ and the two eigenvalues
necessarily acquire opposite signs. Consequently, one perturbation mode
decays exponentially toward the downstream equilibrium while the other grows
exponentially away from it. The downstream state therefore possesses one
stable and one unstable eigendirection, implying that it is a saddle point
of the stationary dynamical system. As a consequence, forward shooting from
the upstream state is numerically unstable, since generic perturbations
excite the forbidden downstream mode and fail to converge to the desired
solution. However, backward integration from the downstream equilibrium is
stable because the forbidden mode becomes exponentially damped in the
reverse direction. This behavior is also observed in the numerical
integrations presented below.

This analysis shows that the asymptotic mode structure remains compatible
with smooth stationary shock solutions as long as the flow remains in the
regular regime. If the regularity condition is not satisfied, the situation
becomes more complicated since it is possible for both eigenvalues to become
positive on the downstream asymptotic states, leading to modes that grow
exponentially away from equilibrium at $\omega \rightarrow \infty $. This
will happen in particular if the downstream velocity satisfies%
\[
v_{f}^{2}-\frac{1}{3}-\frac{2}{3}\frac{\tau _{\eta }}{\tau _{\pi }}>0,
\]%
which happens necessarily once the downstream state exits the regularity sector through $%
v^{2}>v_{+}^{2}$, since it is straightforward to show that $v_{+}^{2}>\frac{1%
}{3}+\frac{2}{3}\frac{\tau _{\eta }}{\tau _{\pi }}$. This is in line with
our initial assumption that the solutions must lie within $\left\vert
v_{-}\right\vert <\left\vert v\left( \omega \right) \right\vert <\left\vert
v_{+}\right\vert $ in order for smooth shock profiles exist.

\subsection{Semi-analytical solutions}

\begin{figure}[]
\begin{minipage}[]
{\linewidth}
\centering
\includegraphics[width=7cm]{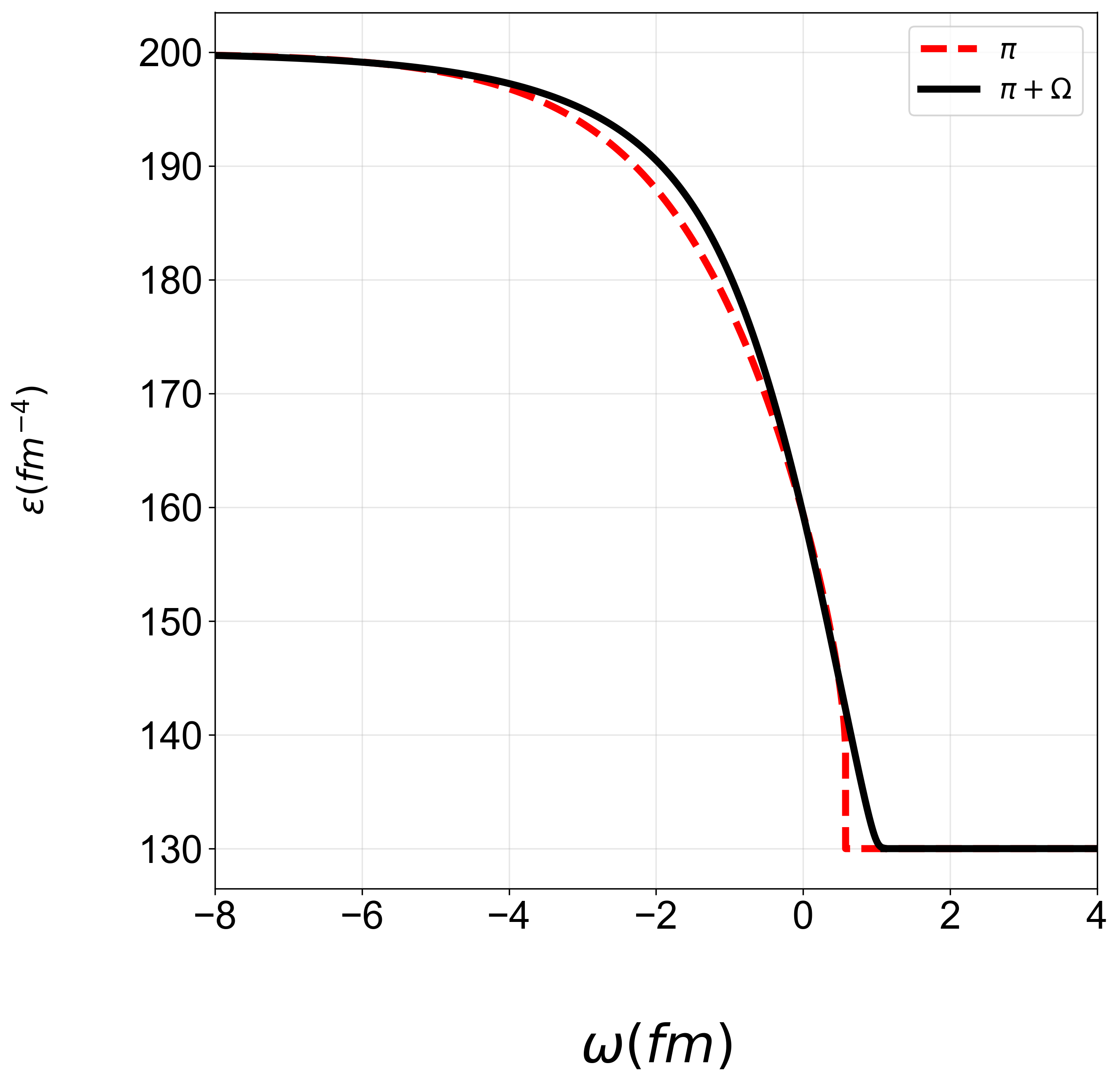}
\end{minipage}
\begin{minipage}[]{\linewidth}
\centering
\includegraphics[width=7cm]
{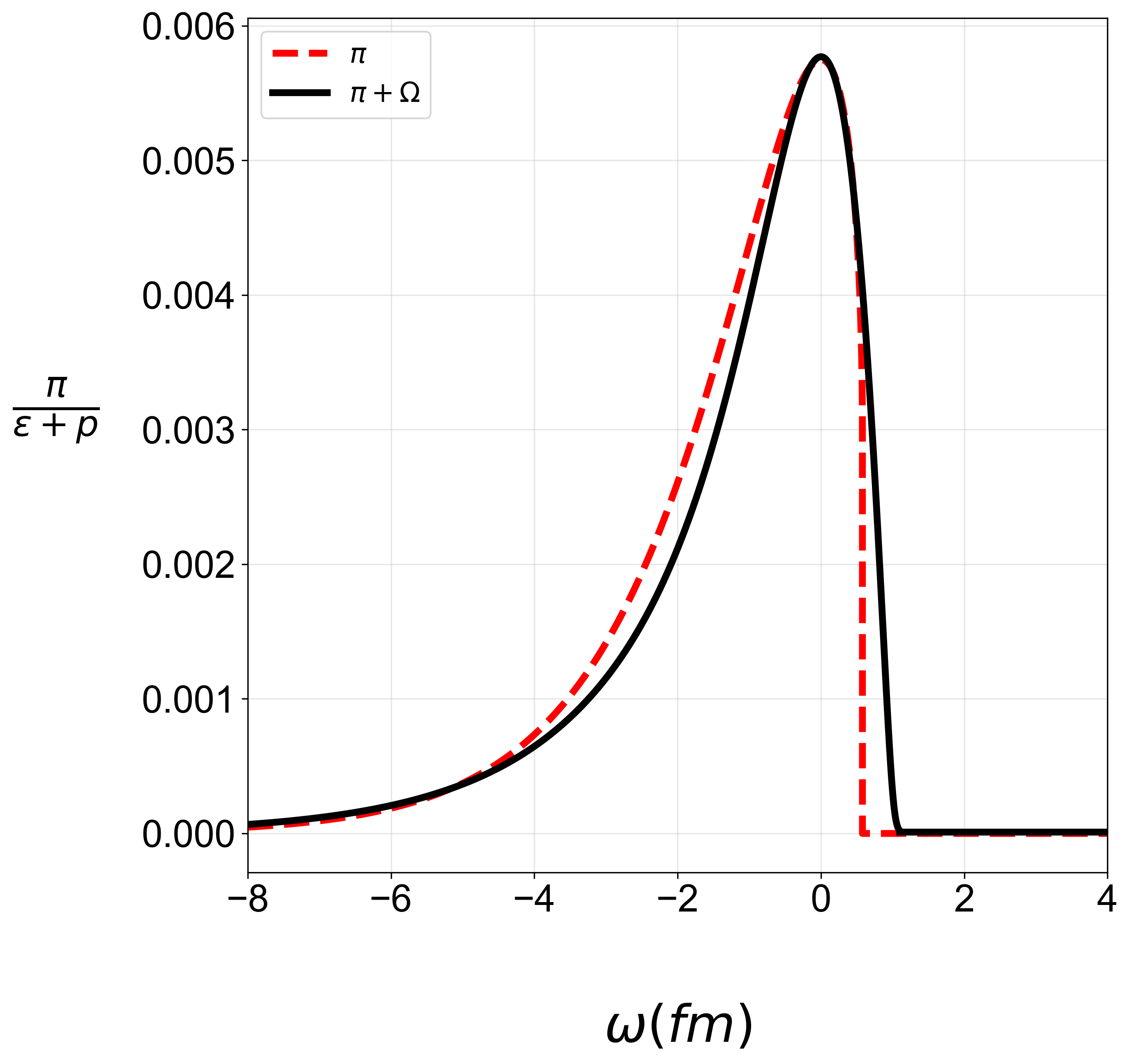}
\end{minipage}
\begin{minipage}[]{\linewidth}
\centering
\includegraphics[width=7cm]{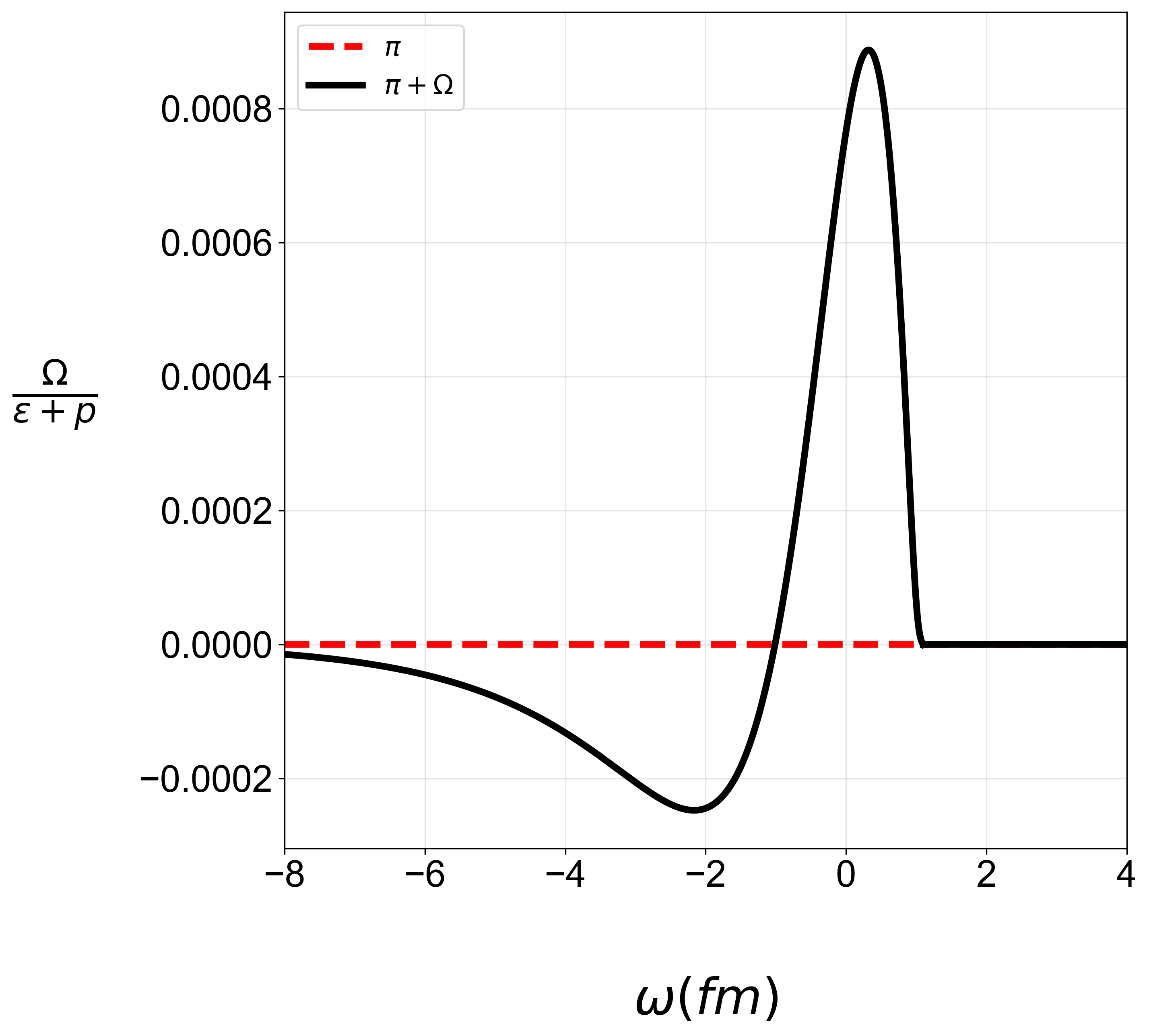}
\end{minipage}
\caption{Comparison between different stationary shock waves, one with only shear viscosity and another with shear and the third order field $\Omega$, where $a_\pi=0.1$ and $b_\pi = 25$.}
\label{fig:Comp_Shear_Heatflow}
\end{figure}
In this section we solve the third-order theory discussed above and determine whether it can regularize certain shock profiles. As before we determine the stationary shock solutions by integrating the stationary equations
using a second-order Runge--Kutta algorithm. We consider the same boundary conditions\footnote{For the red dotted curve we repeat the initial condition provides in Section IV.A}: the constants of motion are $C_{1}=-189.451$ fm$^{-4}$ and $C_{2}=165$ fm, leading to the asymptotic velocities, $v_f=-0.642$ and $v_i=-0.519$, and asymptotic energy densities, $\varepsilon_f=130$ fm$^{-4}$ and $\varepsilon_i=200$ fm$^{-4}$. We then displace the downstream velocity by $v=v_f + 10^{-4}$ and the new degree of freedom by $\Omega = 10^{-4}$ at $\omega _{0}=0.75$ fm, and recalculate the energy density using the relation,

\begin{equation}
   \varepsilon = \frac{C_1}{v} - C_2.
\label{eq:Energia_em_funcao_velocidade}
\end{equation}
The shear-stress tensor component is still calculated using Eq.~\eqref{EquacaoChoqueGrande}. This perturbation from the downstream state will lead to a shock profile, as demonstrated in the previous subsection -- starting with perturbations of the upstream asymptotic state will indeed lead to an instability and no shock profile can be constructed in this form.

\begin{figure}[]
\begin{minipage}{\linewidth}
\includegraphics[width=7cm]{Imagens/ShockVelocity_vs_Vmax_Vmin/Shock_vmax_shear/Choque_Velocidades_Shear_vs_Vmax_dw=0_0001_bulkratio=0_1_tr_shear=25_s1=100_gap=2.png}
\subcaption{$\pi$}
\end{minipage}%
\vspace{0.1cm}
\begin{minipage}{\linewidth}
\includegraphics[width=7cm]{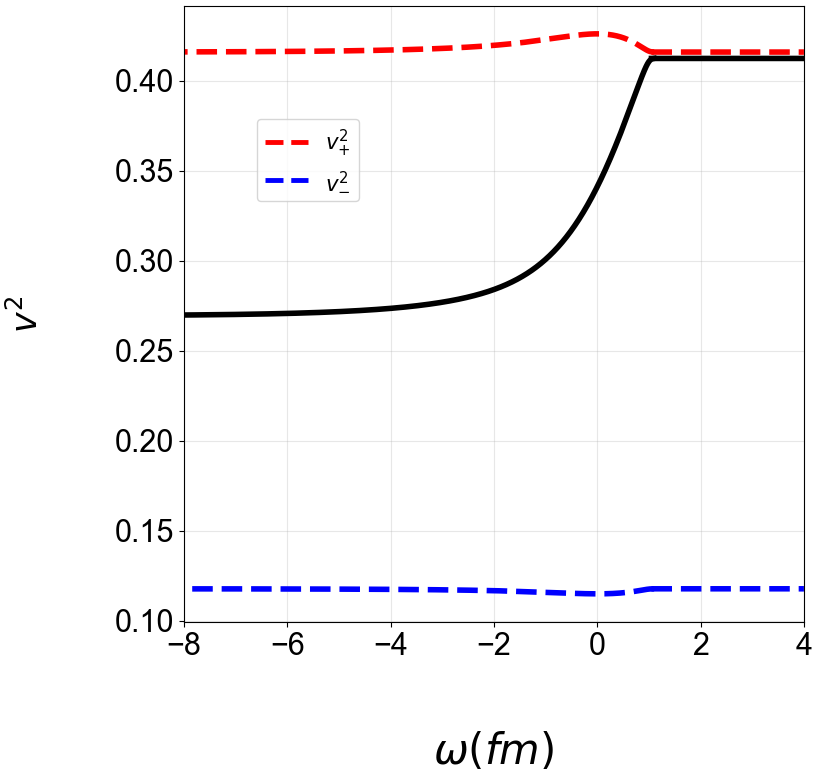}
\subcaption{$\pi + \Omega$}
\end{minipage}
\caption{The velocity and the maximum propagation speed for different stationary shock waves, one with only shear viscosity and another with shear and the third order field $\Omega$, where $a_\pi=0.1$, $b_\pi = 25$.}
\label{fig:Comp_shock_vs_velocidades_min_e_max_Shear_Bulk_Heatflow}
\end{figure}

In Fig. \ref{fig:Comp_Shear_Heatflow} we compare a previous stationary shock solution that displayed singular behavior ($\eta/s = 0.1$ and $b_{\pi}=20$) with a solution of the new theory. The novel transport coefficient is set to $\eta_\Omega = 18/49 \tau_\pi$ and $\tau_\Omega = \tau_\pi$. We see that due to the presence of this new field, the solutions became continuous again. This happened even though the values of $\Omega$ were extremely small, showing that the effect is not due to a significant higher-order contribution to the constitutive relations, but due to the change in the hyperbolic structure of the theory, allowing a wider range of propagation speeds.

We can further check how the new field increases the maximum velocity in Fig.\
\ref{fig:Comp_shock_vs_velocidades_min_e_max_Shear_Bulk_Heatflow}, where the maximum velocity became larger, but a minimum velocity also appeared. Since the velocity profile lies within these two velocity scales, the solution became regular. Nevertheless, this also creates an impediment to completely regularize the theory for all possible shock velocities: if we try to set the maximum (linear) propagation speed to 1, the minimum propagation speed becomes larger leading to loss of regularity near the upstream region. As a matter of fact,   $v_{L-}^2 v_{L+}^2 = \eta_\Omega / (5\tau_\pi)$ and fixing $\eta_\Omega$ to set $v_{L+}^2 \rightarrow 1$ leads to $v_{L-}^2 = 1 / 5$ for $b_\pi=5$. Since the asymptotic upstream velocity becomes $|v_i| \rightarrow 1/3$ when the downstream velocity $|v_f| \rightarrow 1$, the system would certainly become singular in the upstream region.

This result suggests that the description of strong shocks requires the addition of a wide range of nonhydrodynamic fields as one gradually attempts to describe shock profiles with velocities close to the velocity of light. Getting parametrically close to the velocity of light would require an infinite number of fields \cite{Micenmacher1989-fx,Jou1991-od}. This reflects the fact that in relativistic systems the microscopic time scales are dilated in the laboratory frame and small quantities in the local rest frame may become large. In the extreme case of $v\rightarrow1$ this dilation effect becomes infinitely large and no microscopic scale can actually be ignored.

\section{Numerically Regularized theory: Shear and Bulk }
\label{SecVI}

As discussed in the previous sections, the breakdown of smooth shock solutions in Israel-Stewart theory is directly associated with the finite propagation speed of the theory. One possible way to overcome this limitation is to extend the hydrodynamic description by including additional nonhydrodynamic degrees of freedom. In Sec. V, we showed that the inclusion of the third moment of the Boltzmann distribution increases the characteristic propagation speeds and restores smooth stationary shock solutions in situations where Israel-Stewart theory becomes singular. However, the resulting theory contains additional dynamical fields and a more complicated characteristic structure.

If the objective is simply to regularize the equations and extend the range of smooth shock solutions, a much simpler approach is possible. Instead of introducing additional microscopic degrees of freedom, one may directly modify the characteristic propagation speeds through an auxiliary dissipative degree of freedom. In this section we propose such a construction, introducing an artificial causality regulator in the form of an effective bulk-viscous pressure.

Since the fluid under consideration is conformal, this bulk sector is not intended to represent a physical dissipative mechanism. Rather, it serves as a numerical device designed to increase the maximum propagation speed of the theory. Unlike the third-order extension discussed previously, this modification introduces only a single characteristic velocity scale and therefore avoids the appearance of a lower propagation threshold. As a result, it provides a simple framework in which smooth shock solutions can be regularized over the entire range of shock velocities.

This effective bulk viscous pressure enters the energy momentum tensor as a correction to the thermodynamic pressure in the usual way,
\begin{equation}
T^{\mu \nu }=\varepsilon u^{\mu }u^{\nu }-(p+\Pi)\Delta^{\mu\nu
}+\pi ^{\mu \nu }.  \label{TensorEnergiaMomento}
\end{equation}
Then, we must include an equation of motion for this effective bulk viscous pressure that does not spoil the mathematical structure of the theory. We consider a relaxation-type equation of motion a la Israel-Stewart theory \cite
{Denicol:2014vaa}, 
\begin{eqnarray}
\tau_\Pi u^\lambda\partial_\lambda \Pi +\Pi = - \left(\zeta+\frac{2}{3}\tau_\Pi
\Pi\right)\partial_\lambda u ^\lambda,
\end{eqnarray}
where we introduced an effective bulk viscosity, $\zeta$, and an effective bulk relaxation time, $\tau_\Pi$. The equation of motion for the shear-stress tensor remains being given by Eq.~\eqref{IsraelStewartMusic}.

The idea is to consider the bulk viscous pressure to act as a numerical regularization field. So we shall set the bulk viscosity to be extremely small, but the finite ratio $\zeta/\tau_\Pi$ will lead to a correction to the maximum propagation speed that can smooth out the solution without introducing any considerable dissipative effect.

Now, we follow the same steps outlined in the previous sections and reexpress these equations of motion in one dimension for a traveling wave solution. In this case all fields depend solely on $\omega= x-v_{\text{shock}} t$. As before, the conservation laws can be integrated out to obtain the constants of motion, 
\begin{eqnarray}
(\varepsilon+p+\pi + \Pi )\gamma^2v &=& C_1,  \label{Constante1Coupling}
\\
(\varepsilon+p+\pi)\gamma^2v^2+ p + \pi + \Pi &=& C_2,
\label{Constante2Coupling}
\end{eqnarray}
and the equation of motion for bulk and shear pressure become \cite%
{debrito2023thirdorder}, 
\begin{eqnarray}
\tau_\Pi \gamma v \partial_\omega\Pi + \Pi &=& 
-\gamma^3 \left( \zeta+\frac{2}{3}\tau_\Pi\Pi \right) \partial_\omega v, \label{BulkShear1}\\
\tau_\pi \gamma v \partial_\omega \pi +\pi&=& -\frac{
4}{3}\gamma^3\left( \eta+\tau_\pi\pi \right) \partial_\omega v .
\label{BulkShear2}
\end{eqnarray}

Following the same procedure developed in the previous sections, we can manipulate the conservation laws and obtain the following identity,
\begin{equation}
-C_{1}v^{2}+\frac{4}{3}C_{2}v-\frac{C_{1}}{3}=(\pi+\Pi) v.
\label{ChoqueGrande2}
\end{equation}
This equation is similar to \eqref{EquacaoChoqueGrande} obtained in the presence of just shear viscosity. Taking the derivative of Eq.~\eqref{ChoqueGrande2} and using the equations of motion for the dissipative currents, Eqs.~\eqref{BulkShear1} and \eqref{BulkShear2}, we obtain the fundamental ordinary differential equation satisfied by the velocity field,
\begin{equation}
\left( \varepsilon +p+\pi+\Pi \right) \left( v^{2}-v_{\mathrm{SB}}^{2}\right)
\gamma^{3}\partial _{x}v=\frac{\pi }{\tau_{\pi}}+\frac{\Pi}{\tau_{\Pi }}.  \label{MasterEquation2}
\end{equation}
The addition of a bulk viscous pressure modifies it by introducing corrections to the thermodynamic pressure and the dissipative source term on the right hand side, and, most importantly, by modifying the maximum propagation speed, which now becomes,  
\begin{equation}
v^2_{\mathrm{SB}} = \frac{1}{3} + \frac{1}{\varepsilon+p+\Pi+\pi}\left(\frac{\zeta}{\tau_\Pi} + \frac{4}{3}\frac{\eta}{\tau_\pi } + \frac{2}{3}\Pi + \frac{4}{3}\pi \right).
\label{blabla}
\end{equation}
Here we can see how the inclusion of a bulk viscous pressure can lead to significant modifications to the solutions of the hydrodynamic theory. While the bulk viscosity can be very small and lead to negligible corrections to the thermodynamic pressure, the ratio $\zeta/\tau_\Pi$ does not have to be small and we can use it to significantly increase the characteristic speed of the theory and smooth out the shock profile. In contrast to the novel theory described in the previous section, this procedure does not introduce another propagation speed. 

The equation is regular as long as,
\begin{equation}
    v_{\textrm{shock}} \neq v_{\mathrm{SB}}.
\end{equation}
Nevertheless, given the boundary conditions considered for the solutions with only shear viscosity, which start at $|v|\approx|v_f|<|v_{\mathrm{SB}}|$, and assuming that the velocity profile remains monotonic, a sufficient condition to obtain regular solutions is that, 
\begin{equation}
    v_{\textrm{shock}} < v_{\mathrm{SB}}.
\end{equation}
Here, since we treat the bulk relaxation time as a free parameter of the theory, we can always choose it in such a way that $v_{\mathrm{SB}}\approx 1$ and, thus, this condition is always satisfied. We shall confirm this by solving the equations in the following section.

\subsection{Semi-analytical solutions}
\begin{figure}[]
\begin{minipage}{\linewidth}
\centering
\includegraphics[width=7cm]{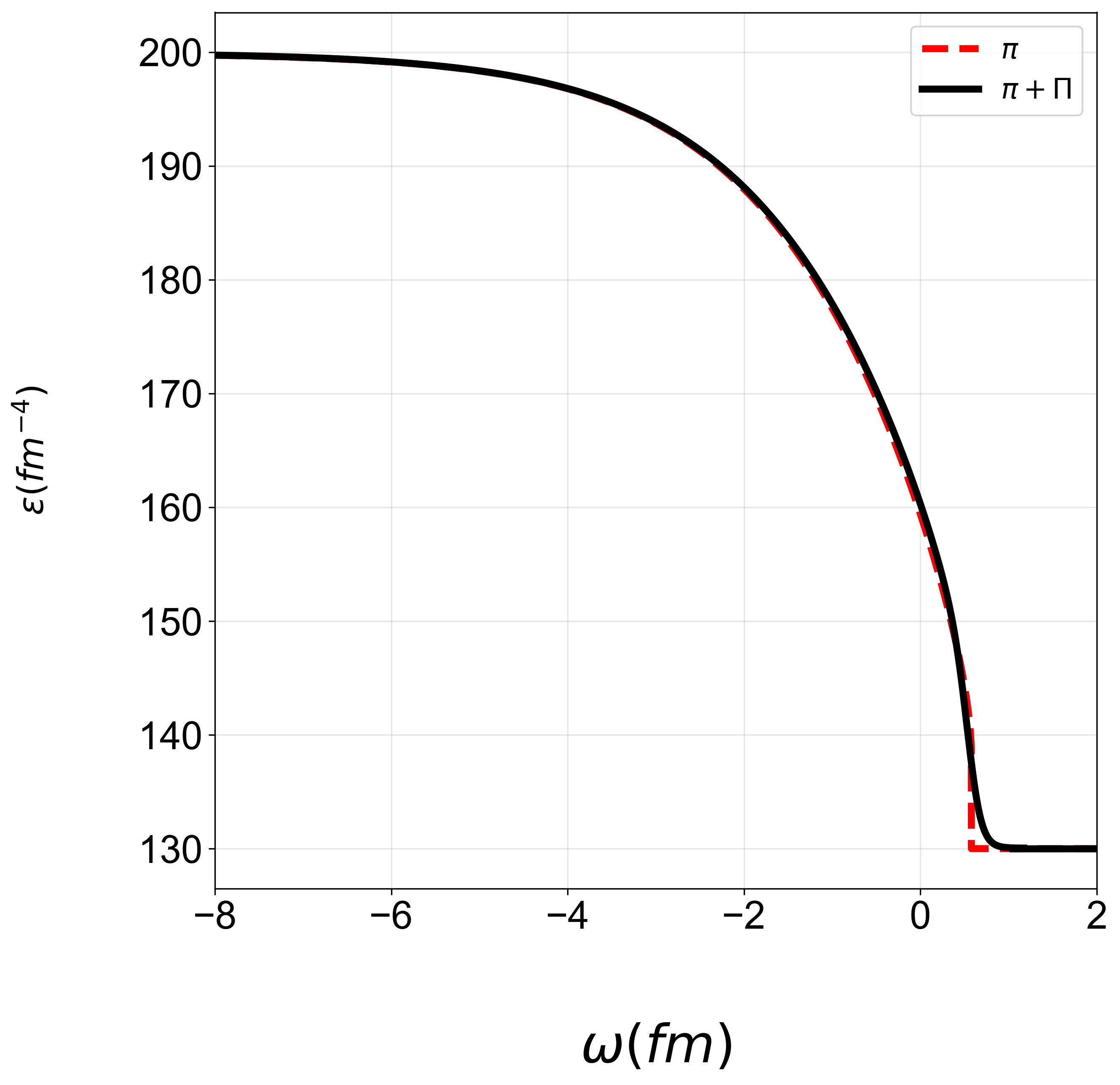}
\end{minipage}
\begin{minipage}{\linewidth}
\centering
\includegraphics[width=7cm]{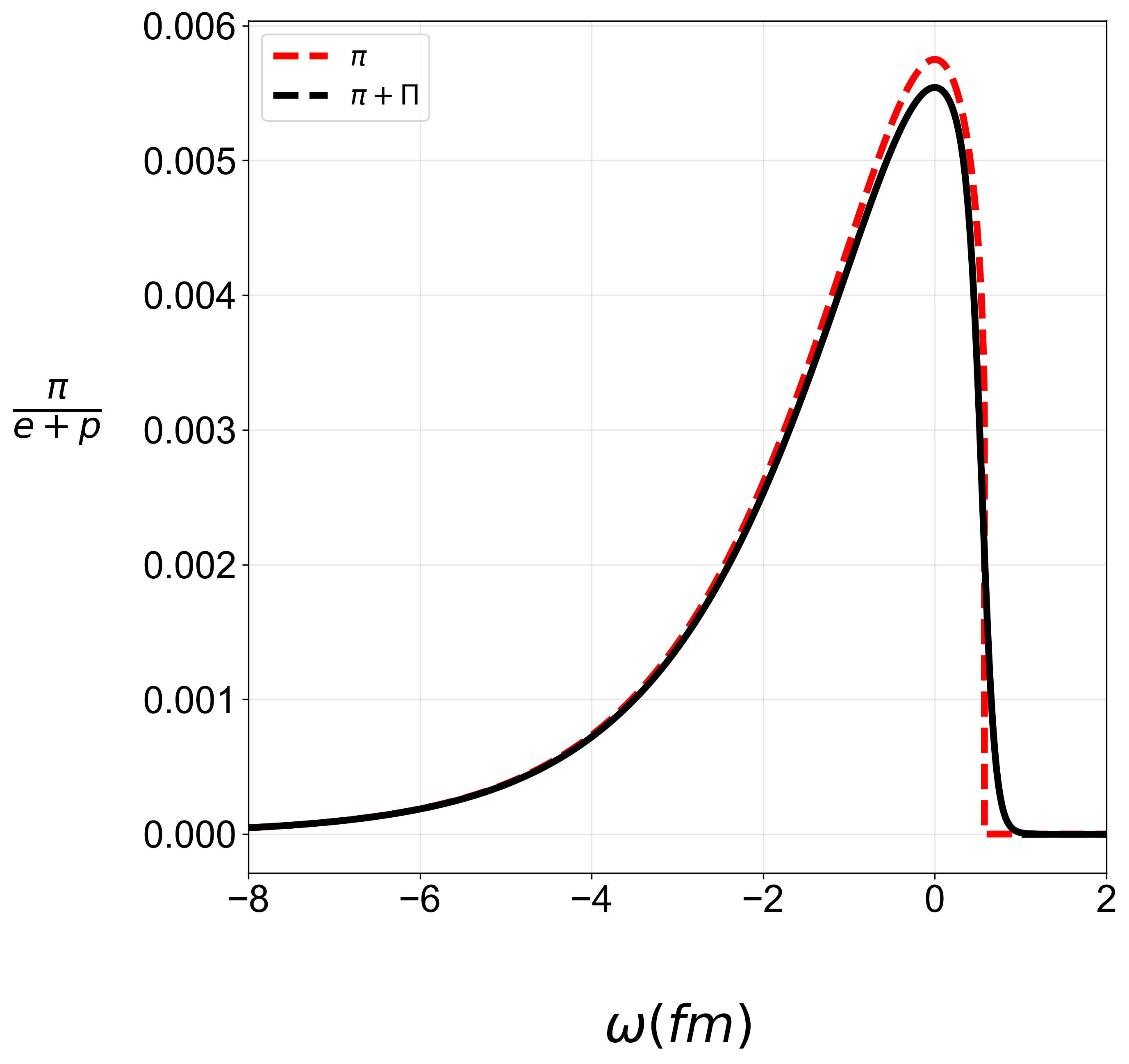}
\end{minipage}
\begin{minipage}{\linewidth}
\centering
\includegraphics[width=7cm]{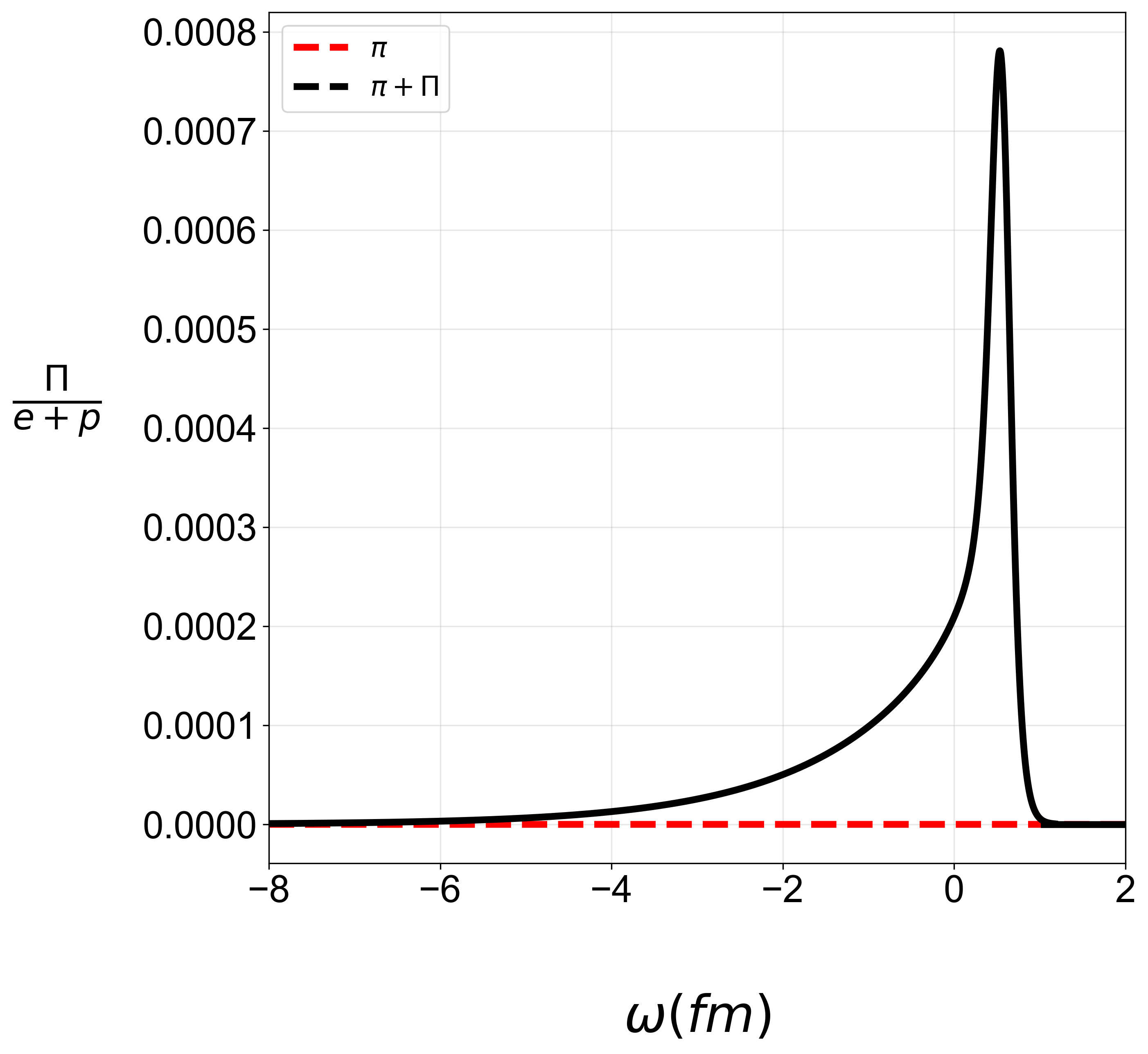}
\end{minipage}
\caption{Comparison between different stationary shock waves, one with only shear viscosity and another with shear and bulk, where $a_\pi=0.1$, $b_\pi = 25$, $a_\Pi = 0.005$ and $b_\Pi=5$ 
}
\label{imagem: Comp Shock-Vs-Shock+Bulk}
\end{figure}

\begin{figure}[]
\begin{minipage}
{\linewidth}
\includegraphics[width=7cm]{Imagens/ShockVelocity_vs_Vmax_Vmin/Shock_vmax_shear/Choque_Velocidades_Shear_vs_Vmax_dw=0_0001_bulkratio=0_1_tr_shear=25_s1=100_gap=2.png}
\subcaption{$\pi$}
\end{minipage}
\begin{minipage}{\linewidth}
\includegraphics[width=7cm]{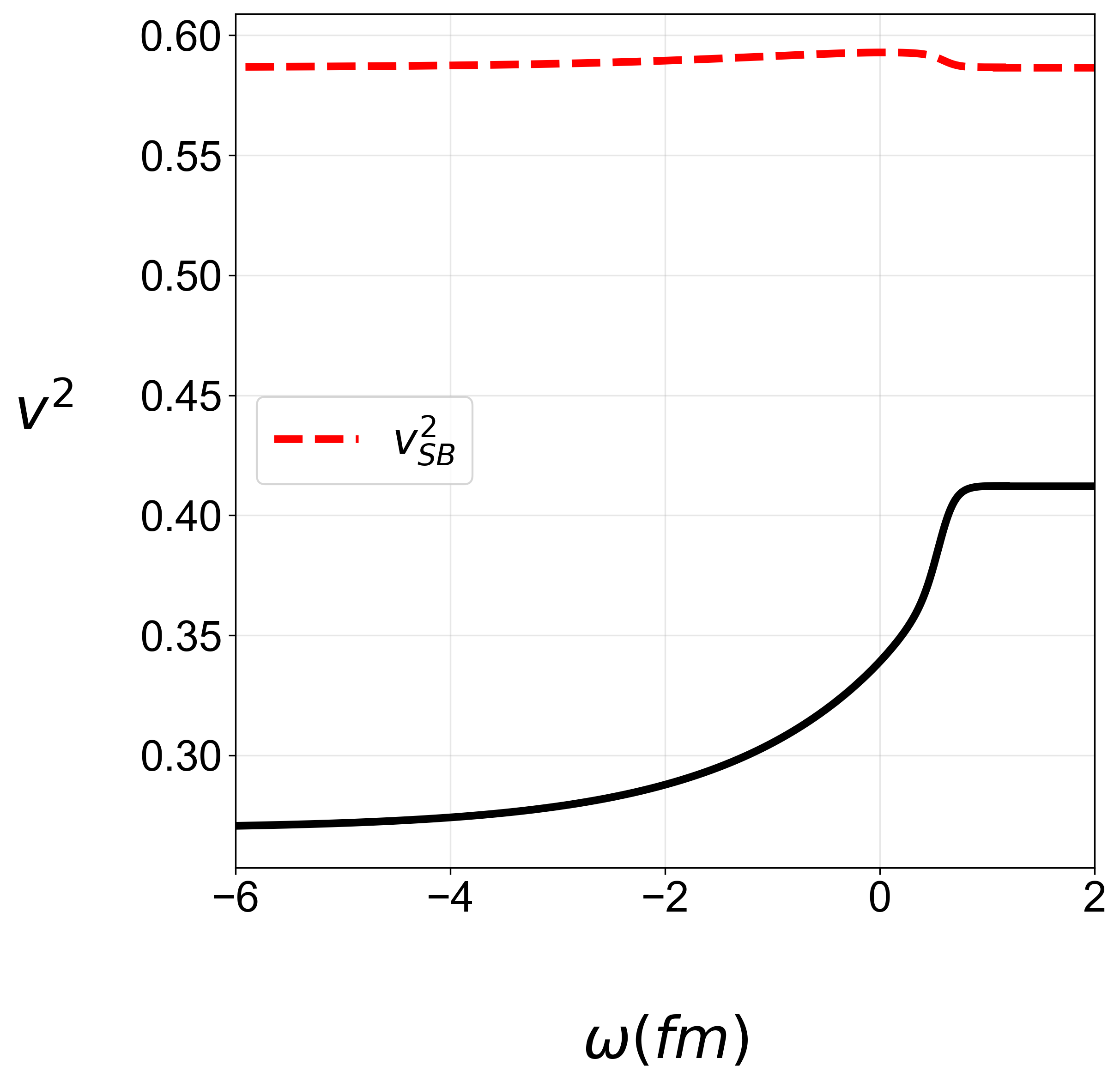}
\subcaption{$\pi + \Pi$}
\end{minipage}%
\caption{The velocity and the maximum propagation speed for different stationary shock waves, one with only shear viscosity and another with shear and bulk, where $a_\pi=0.1$, $b_\pi = 25$, $a_\Pi = 0.005$ and $b_\Pi=5$ and the maximum velocity for each dissipation.}
\label{fig:Comp_shock_vs_velocidades_min_e_max_Shear_Bulk}
\end{figure}
Once again, we solve the Israel-Stewart theory, now including an effective bulk viscous pressure, and determine whether it can regularize certain shock profiles. As before we determine the stationary shock solutions by integrating the stationary equations
using a second-order Runge--Kutta algorithm. We consider the same boundary conditions\footnote{For the red dotted curve we repeat the initial condition provided in Section IV.A.}: the constants of motion are $C_{1}=-189.451$ fm$^{-4}$ and $C_{2}=165$ fm, leading to the asymptotic velocities, $v_f=-0.642$ and $v_i=-0.519$, and asymptotic energy densities, $\varepsilon_f=130$ fm$^{-4}$ and $\varepsilon_i=200$ fm$^{-4}$. As before, we displace the downstream velocity by $v=v_f + 10^{-4}$, and we perturb the dissipative fields by a small positive value, $\pi =\Pi = 10^{-4}$ at $\omega _{0}=0.75$ fm, and recalculate the energy density using the relation \eqref{eq:Energia_em_funcao_velocidade}. This perturbation from the downstream state will lead to a shock profile
 \footnote{Similar to the case studied in the third-order hydrodynamics, starting with perturbations of the upstream asymptotic state will lead to an instability and no shock profile can be constructed in this form. This can be demonstrated by a linear stability analysis that we will not develop here.}.
The shear and bulk relaxation time will be parametrized in the following way \cite{Denicol:2014vaa},
\begin{eqnarray}
        \tau_\Pi = b_\Pi \frac{\zeta}{\varepsilon+p} \,\,\,\,\,\text{and} \,\,\,\,\, 
        \tau_\pi = b_\pi\frac{4\eta}{3(\varepsilon+p)}. 
    \label{eq: tau_Pi e tau_pi}
\end{eqnarray}
This parametrization ensures that the theory can be causal and stable. The bulk viscosity is assumed to be proportional to the entropy density, as was assumed for the shear viscosity, $\zeta = a_\Pi s$. We shall only consider small values of $a_\Pi$ so that the dominant effect of this effective bulk viscous pressure is to increase the maximum propagation speed of the theory. 

Finally, we can compare two solutions, one with both viscosities and another
with only shear viscosity, in Fig.\ \ref{imagem: Comp Shock-Vs-Shock+Bulk}. It is clear that the addition of the bulk pressure makes the previous solutions continuous again: this is a consequence of increasing the maximum propagation speed of the theory, going from $v^2_{NL}$ to $v^2_{SB}$. This can be clearly observed in Fig.\ \ref{fig:Comp_shock_vs_velocidades_min_e_max_Shear_Bulk} where we compare the maximum velocity of each case. In fact, the presence of bulk viscosity increases the range of velocities that lead to a continuous shock solution, since we can increase the maximum propagation velocity without any drawback related to a minimum propagation velocity. Although the inclusion of bulk viscosity is not physical, it is only a numerical regulator. This result emphasizes that the breakdown of continuous shock solutions is an information transmission problem caused by the limited propagation speed in Israel-Stewart theory.

Both the numerical bulk regulator and the third-order extension presented in Section V share a common conclusion: the subshock structure is a consequence of the finite propagation speed in Israel-Stewart theory, not a genuine physical effect. This may indicate a fundamental limitation of Israel-Stewart theory, which may not reliably describe shock waves in the ultra-relativistic limit.

\section{Conclusions}
In this work, we investigated stationary shock waves in conformal Israel-Stewart theory and established the conditions under which smooth shock profiles exist. We showed that smooth stationary shock profiles exist only while the shock velocity remains below the maximum characteristic propagation speed supported by the nonlinear hydrodynamic equations. Beyond this threshold, the stationary shock equations become singular and no continuous solution can connect the upstream and downstream equilibrium states. The shock must therefore be completed by a discontinuity satisfying the Rankine-Hugoniot conditions. The analytical predictions obtained from the stationary equations were corroborated through numerical simulations of the relativistic Riemann problem.

An important outcome of this work is that the breakdown of smooth shock solutions is controlled by the characteristic structure of the hydrodynamic equations rather than by the magnitude of the dissipative corrections. In all cases considered here, the dissipative stresses remain finite as the stationary equations become singular. The loss of regularity therefore reflects the inability of the truncated hydrodynamic theory to propagate information across the shock profile sufficiently rapidly to sustain a continuous solution, rather than the conventional breakdown of hydrodynamics associated with large departures from local equilibrium.

To further investigate this interpretation, we considered two distinct regularization strategies. The first extends Israel-Stewart theory by promoting the lowest neglected kinetic moment to an independent dynamical field, thereby modifying the characteristic propagation speeds of the theory. While this enlarges the range of smooth shock solutions, it also introduces a second characteristic velocity that ultimately limits the achievable extension of the regular regime. The second introduces a small effective bulk-viscous pressure as a numerical regulator. Although this modification is not intended to represent a physical dissipative mechanism in conformal fluids, it increases the maximum propagation speed without introducing additional characteristic constraints and successfully restores smooth stationary shock solutions. Despite their different constructions, both regularization strategies point to the same physical conclusion: the existence of smooth shock profiles is governed by the characteristic propagation structure of the hydrodynamic theory.

Our results establish a direct connection between the characteristic structure of causal relativistic hydrodynamics and the existence of smooth stationary shock solutions. They suggest that the applicability of truncated dissipative theories is constrained not only by the magnitude of dissipative corrections, but also by their ability to propagate information at speeds compatible with the underlying nonlinear flow. Whether the composite shock structures identified here survive in more microscopic descriptions, such as solutions of the Boltzmann equation or other kinetic theories, remains an important open question.


\section*{Acknowledgments}

The authors thank Jorge Noronha and Douglas S.~N.~Domingues for fruitful discussions. G.~S.~D.~is supported by CNPq through the grant 307761/2022-3. D.~D.~is partly funded by CAPES. G.~S.~D.~and D.~D.~acknowledge the support of CNPq and FAPERJ via the INCT-FNA grant 408419/2024-5.

\bibliographystyle{unsrt}
\bibliography{ref}
\end{document}